\documentclass{article}
\usepackage{graphicx, comment, jr,color,multicol,enumerate}

\def\c14{${}^{14}$C}
\def\bp{\fun{BP}}

\begin{document}

\title{The utility of a  Bayesian analysis of complex models and the study of archeological \c14 data.
}

\author{\jr\\Department of Statistics, The Hebrew University of Jerusalem\\
\& Department of Statistics, University of Michigan, Ann Arbor\footnote{Department of Statistics,
University of Michigan,
1085 South University,
Ann Arbor, MI 48109-1107; yritov{$@$}umich.edu}
}

\maketitle
\begin{abstract}
The paper presents a critical introduction to the complex statistical models used in \c14 dating. The emphasis is on the estimation of the transit time between a sequence of archeological layers. Although a frequentist estimation of the parameters is relatively simple, confidence intervals constructions are  not standard as the models are not  regular.  I argue that that the Bayesian
paradigm is  a natural approach to these models. It is simple, and gives immediate solutions to credible sets,
with natural interpretation and simple construction. Indeed it is the standard tool of \c14 analysis. However and   necessarily, the Bayesian approach is based on technical  assumptions that may dominate the scientific conclusion in a hard to predict way.  I  exemplify the discussion in two ways. Firstly, I simulate toy models.  Secondly, I analyze  a particular data set from the Iron Age period in Tel Rehov. These data  are important to the debate on the absolute time of the Iron Age I/IIA transition in the Levant, and in particular to the feasibility of the Bible story about the United Monarchy of David and Solomon. Our conclusion is that the data in question cannot  resolve this debate.
\end{abstract}

\section{Introduction}

Statistical inference is built on the assumption that the observed data tells us  about the parameters
of interest. Typically, this is translated into a statement that the data are random, and their distribution
function depends on real unknown parameters. In the context of radiocarbon data, this is, for example, the statement that the
laboratory carbon dating is well approximated by a normal
random variable with unknown mean and unknown variance. The term random variable here is understood in the sense
of a counterfactual. If the lab conducts the same experiment again and again, with similar specimens, the result of each experiment will be different, with actual distribution as prescribed by the statistical model.
The important thing is that to a large extent this (statistical) model is objective, based on the physics of the
experiment, and can be verified by empirical study (i.e., repeating the measurements).  In
fact, the sort of measurements used in radiocarbon dating have a built-in repetition mechanism, e.g., the days in
counting measurements.

The Bayesian paradigm builds over this another layer of randomness, where the unknown parameters themselves
are considered random, but this randomness has a different meaning than that of the measurement random error. It
is a way to describe what the data analyzer thought and knew about these parameters before observing the data. It
is the \emph{a priori} knowledge on the subject matter. This information is summarized in the \emph{a priori}
distribution. The Bayesian approach may be not the dominant approach to applied statistics, but it is the
ruling one in radiocarbon archaeological dating (Steier and Rom 2000; Manning 2001; Bayliss and Bronk Ramsey 2004;  Bronk Ramsey 2009). The standard statistical packages used in archaeology are Bayesian:
The OxCal (Bronk Ramsey and Lee 2013), and the BCal (Buck, Christen and James 1999).

The first randomness can be simple, justifiable, and safe.   However, the second, the \emph{a priori} distribution, typically lacks objectivity,
describes a unique situation, and cannot be verified experimentally.

Whatever the data analysis is, the researcher should be careful to ensure that the analysis is robust in the
sense that the conclusions are derived from data and not from non-valid assumptions (Berger and Berliner 1986;
Berger 1985 ch. 4.7).
In the context of radiocarbon dating of archeological sites, distribution assumptions enter in many ways, In
particular we should have
\begin{enumerate}
\item
	Assumptions about the distribution of the laboratory measurements error (typically and for convenience
assumed to be normal).
\item \label{ass2}
	Assumptions about the calibration curves used (see below).
\item
Assumptions about the archeological context, stratigraphic sequences, and the time order of different findings.\label{arch}
\item
	Implicit \emph{a priori} assumptions on the dates involved\label{ass3} given the archeological constraints.
\end{enumerate}
The conceptually difficult assumptions are those of point \ref{ass3}, and similarly, but to less extent, those of point
\ref{ass2}. They are done as a matter of fact without much regard to their implications. Typically, they are just
part of the algorithm (e.g., using OxCal or BCal in archeology), and the subject matter scientist is not aware of them.

In Ritov et al.\ (2014) we  argued that Bayesian priors for complex models should be chosen with special care. They may lead to biased estimators in unpredicted ways. Although
it is usually true that good estimators can be based on well chosen priors, the prior can be justified only by sophisticate
investigation of the theoretical behavior of the resulting estimators. Moreover, we argued that the fact that a
prior formalizes  reasonable assumptions on reality is not enough, and a plausible prior may lead to a counter-intuitive bias. A good prior for one scientific question (e.g., the starting time of a period),
may yield a bad answer for a different question (e.g., its length), even if both answers are based
on the same data.    In the current paper, I try to investigate the impact of prior assumptions on the analysis of the \c14
complex data from archeological strata.

The Bayesian approach is a simple way to introduce any theory the researcher has in addition to the
radiocarbon data. The results of the analysis presented by a Bayesian researcher do not differentiate between this theory and hard evidence.  Any theory used in the analysis modifies the conclusions, otherwise there was
no rationale to introduce it.   The dating of a layer by an archeologist who believes  the Biblical text is reliable,  should be different from the conclusions of the analysis of the same data by a researcher who believes that story of the United Monarchy (of David and Solomon) was invented by a later regime. Of course, this is not the way the Bayesian paradigm is used in practice, but then, I believe, it loses its philosophical grounds.

The structure of this paper is the following.   The impact of different statistical approaches to the the analysis of this type of data is exemplify in Section  \ref{sec:toys}. This is done by the presentation of  simple models and their simulations. In Section \ref{sec:calibration},  a somewhat naive introduction to the analysis of the  calibration curve is given, and it is argued that the standard Bayesian approach may be problematic. In Section \ref{sec:rehovIntroduction}, I present the data to be analyzed later, that of Tel Rehov, where a new parallel analysis of the data using Bayesian and non-Bayesian methods is given, and in particular we argue that a standard statistical approach may dictate a simpler analysis of the calibration curve than is used in practice. Finally some concluding remarks are given in Section \ref{sec:summary}.

\section{Archeological Bayesian analysis: simulations}
\label{sec:toys}

In this section I will consider a sequence of three simple and artificial examples which cover the ingredient of the analysis of the data from Tel Rehov starting with almost the simplest possible model. In all the cases I would check to what extent the conclusions are sensitive to the technical
assumptions.   I use simulations, thus it is easy to evaluate the performance
of the estimators as the truth is knwon.

\subsection{A transition between two periods}

In a typical situation data is collected from two consecutive strata. Findings from the two strata are dated. The
real archaeological question may be the time of the transition between the two periods represented by the strata.
In the following I exemplify that sometimes data that are  irrelevant to the archaeological question may
influence quite heavily the Bayesian conclusion.

Consider the following example. Stratum I started at the known time $t_s=1100$ BCE and ended
at the unknown time $\tau$. Stratum II started at $\tau$  and ended at the known time $t_e=900$ BCE. We have two observations at time 1000 BCE, one from each stratum.  From stratum I there were also K-1 measurements from  sources originated at time $1100$, while from stratum II we had M-1
measurements from around time $900$. All measurements were with standard error equal
to 10 years.  We want to estimate the transition time $\tau$.

A frequentist would say that given the two data points at $1000$, $\tau$ should be somewhere between $1020$
to $980$  BCE,  as it should be no more than two standard deviations from these points. But if so, the $K+M-2$
remote points are at least three standard deviation from $\tau$ and hence tell us almost nothing about its value.
By the symmetry of  the likelihood, he would estimate $\tau$ by $1000$ and the confidence interval should be symmetric around this value.

\begin{figure}[t]
\begin{center}
\includegraphics[width=1\textwidth]{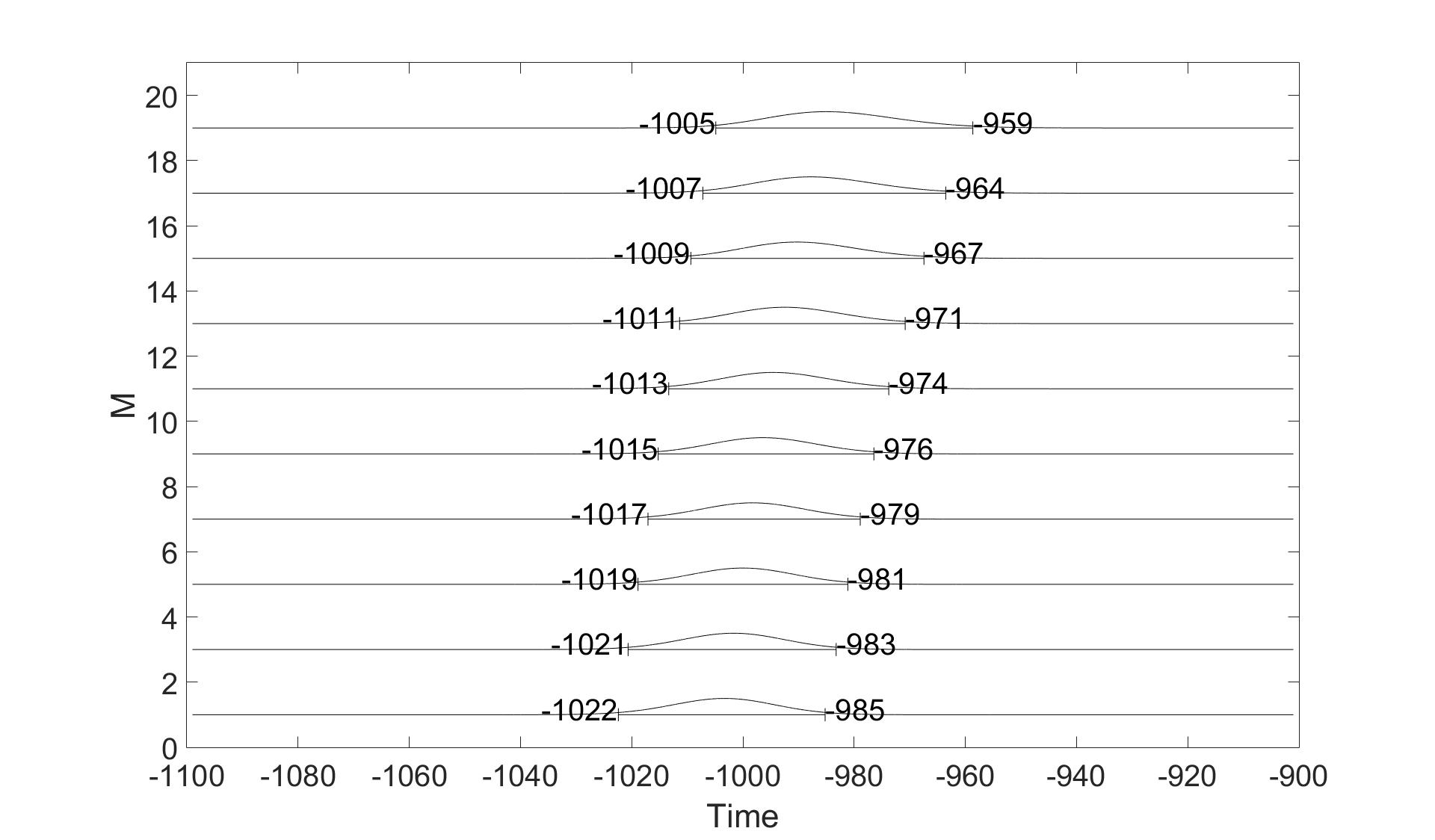}
\caption{Bayesian credible intervals for the time of a single transition as a function of the number of remote
observations, each interval fits different value of $M$, see the text for details.}
\label{BCIremote}
\end{center}
\end{figure}

Consider now the Bayesian analysis. It seems natural to assume that $\tau$ has \emph{a priori}  a uniform distribution on the interval $(t_s,t_e)$, and
given $\tau$,  the distribution of the particular dates would be independent and uniform on $(t_s,\tau )$ and
$(\tau ,t_e)$ respectively. On the other hand, it seems reasonable to assume that the estimate of $\tau$  should
be based only on the two measurements at $1000$ BCE. The rest of the measurements seem to be irrelevant to the dating of $\tau$. They are
simply too remote from the boundary. However, this is not so.
In Figure \ref{BCIremote}, I plotted the \emph{a posteriori} credible intervals as a function of M, for  $K=5$. The  remote observations influence  dramatically the credible sets is  because the uniform distribution on $(t_s,\tau )$
has density which is equal to $1/(\tau -t_s )$. Since we have $K$ observations from this interval we get a factor
of $(\tau -t_s )^K$, i.e., the total number of observations , $K$, from stratum I has an influence on the
Bayesian conclusion. Similarly, there is a factor of $(t_e-\tau)^M$ factor due to Stratum II. The latter factor favors short Stratum II time, and thus as number of remote young findings increases, the \emph{a posteriori} distribution tends to shorter second periods.

Now, the flat prior of $\tau$ can be corrected  to balance out these factors in the likelihood, i.e. we could use a
prior which is proportional to $(\tau -t_s)^K (t_e-\tau )^M$. This would yield a reasonable solution to the
specific findings with the specific values of $M$ and $K$, at the price of a very informative prior, which would
not fit any other findings in this area. Roughly speaking, any admissible statistical procedure can be
constructed using some priors, but they may be too ad hock, and should fit exactly the particular scientific
questions asked. In particular, answers to different scientific questions should be based on different priors
even if all of them are based on the same lab measurements  (Ritov et al. 2014).

\begin{figure}[t]
\begin{center}
  \includegraphics[width=0.495\textwidth]{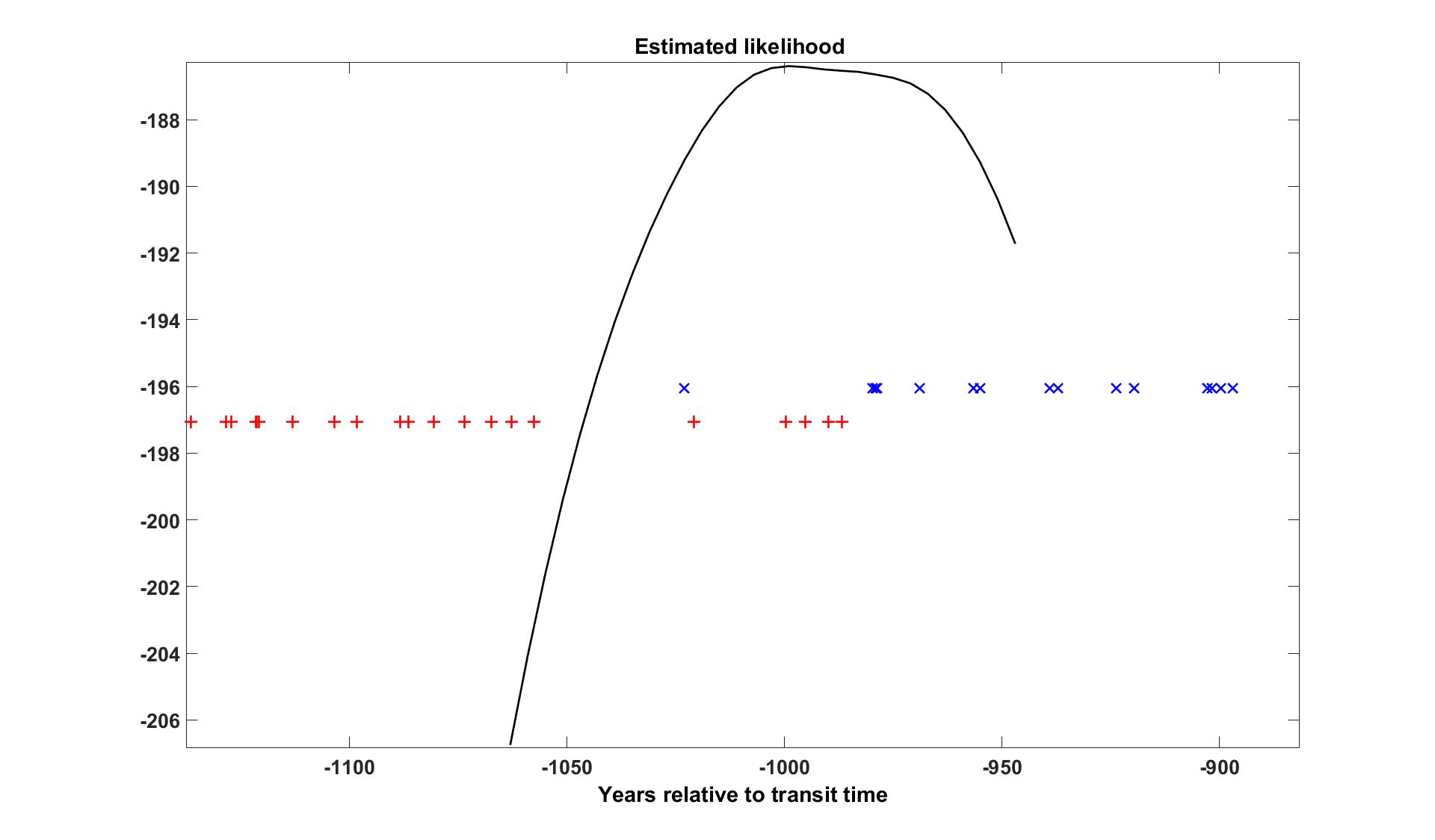}
  \includegraphics[width=0.495\textwidth]{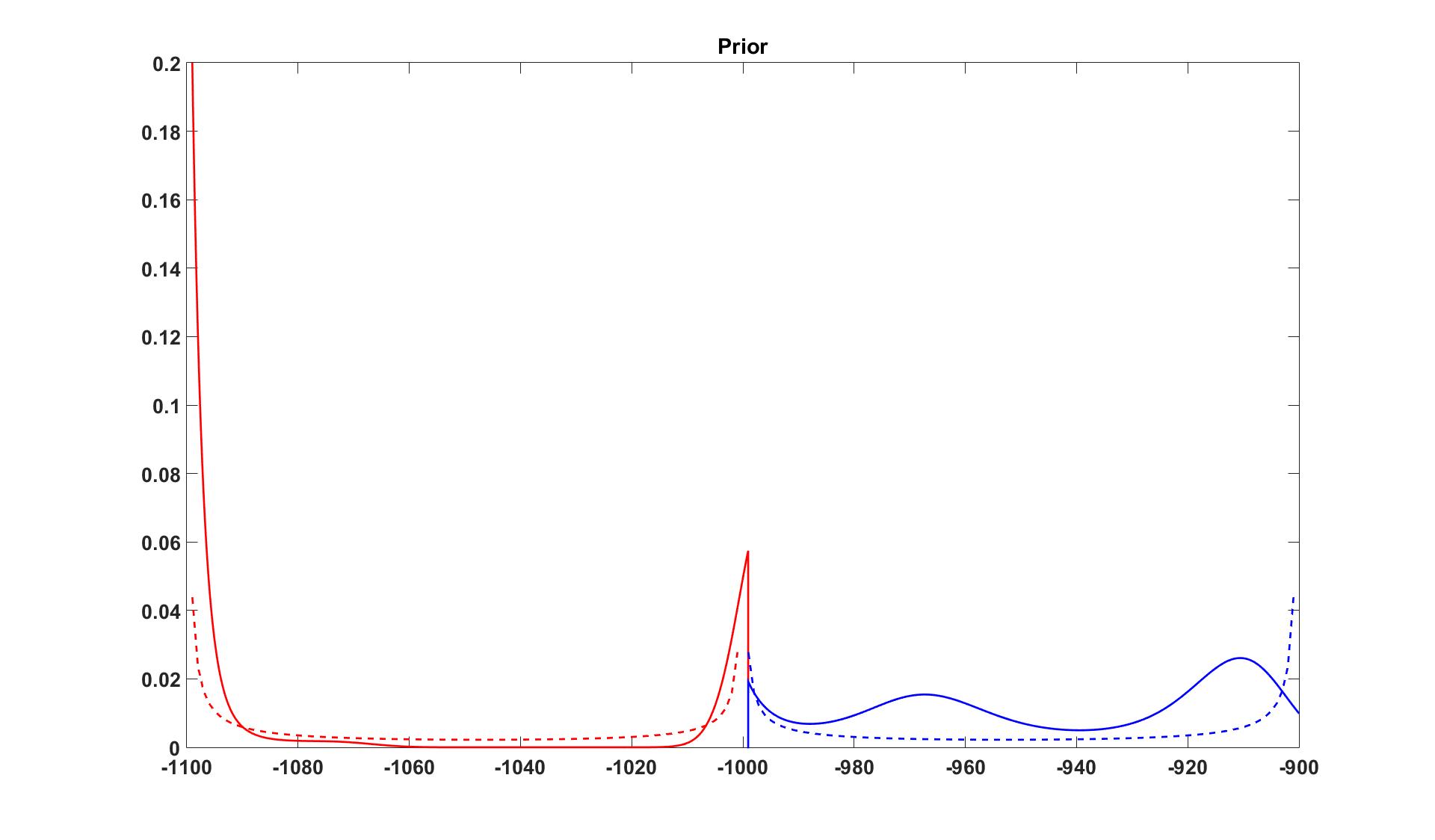}
  \caption{Empirical Bayes estimator of the change time. The true prior distributions are $\pi_1 = Be(0.1,0.2)$ and $\pi_2 = Be(0.2,0.1)$ (the broken lines on the right figure). The estimators of $\hat\pi_i$ (solid lines) were MLE on a grid using EM algorithm with an early stop. The observations of one sample are plotted on the left figure ($+$ for the first layer and $x$ for the second) as well as the likelihood function. }
  \label{fig:eb}
\end{center}
\end{figure}

There is an intermediate way which is the \emph{empirical Bayes} approach. In this approach we can assume that $Y_{ij}\dist N(\tau_{ij},\sig_{ij}^2)$, and $\tau_{ij}\dist \pi_i$, $i=1,2$, $j=1,\dots,n_i$, where $\sig_{ij}$ are known,  $\tau_{ij}$ and $\pi_i$ are unknown except that $\supp\pi_1\subseteq [t_s,\tau]$ and $\supp\pi_2=[\tau,t_e]$. Statistically, deconvolution is difficult, and hence  $\pi_1$ and $\pi_2$ can be estimated only in a very slow rate. However,  estimating of the smooth $\pi_i*N(0,\sig_{ij}^2)$ is easy, and estimation of the Bayes procedure is again an easy task. In Figure \ref{fig:eb} we presents an analysis of the example. The distributions $\pi_1$ and $\pi_2$ where estimating by (an approximate) nonparametric MLE (on a grid) using the EM algorithm.

\subsection{A sequence of events}
We consider now the timing of a sequence of events $\mu_1,\mu_2,\dots,\mu_M$, and we suppose that we have
independent radiocarbon dating of these events, $Y_1,Y_2,\dots,Y_M$. We assume that $Y_m$ is Gaussian with mean
$\mu_m$ and known standard deviation $\sig_m$. We assume that it is well established that the times are ordered
such that $\mu_1<\mu_2<\dots<\mu_M$. For simplicity, we add that it is known that $t_s<\mu_1$ and $\mu_M<t_e$. To
simplify notation, we define $\mu_0=t_s$ and $\mu_{M+1}=t_e$.

There is no simple non-Bayesian solution for this problem. Finding the maximum likelihood estimator, the values
of the parameters that maximize the value of the joint probability density at the observations, and obeying the
order restriction is not difficult, although the solution has no close form and should be found by a simple
numerical algorithm. However, the construction of confidence intervals is not simple, e.g., because the model is
not a regular parametric model and the distribution depends heavily on the values of the parameters. Constructing
confidence intervals which do not use the order restrictions is easy, but seems to be inefficient use of the
data. Bayesian estimation and credible intervals seem to be just the right solution. Of course, the price is a
strong dependency on implicit assumptions built into the prior.

It may seem natural to use ``non-informative'' prior, that is, to assume that if $\mu_{m-1}$ and $\mu_{m+1}$ are
known, then $\mu_m$ is uniform in the interval $(\mu_{m-1},\mu_{m+1})$.  However, the order restriction is quite
tight. This prior prescribes that \emph{a priori} $(\mu_m-t_s)/(t_e-t_s)$  is a beta random variable with
parameters $\alpha=m$ and $\beta=M-m+1$. In particular, its mean is $t_s+\frac{m}{M+1}(t_e-t_s)$ and its standard
deviation is $\frac{t_e-t_s}{M+1} \sqrt{\frac{m(M-m+1)}{M+2}} $ . If $t_s=3150 BP$, $t_e=2850 BP$, $M=10$, and
$m=2$ then \emph{a priori} it is assumed that with probability of 0.95, $t_2$  is between 3142 to 3016 BP.

\begin{figure}[t]
\begin{center}
\includegraphics[width=1\textwidth]{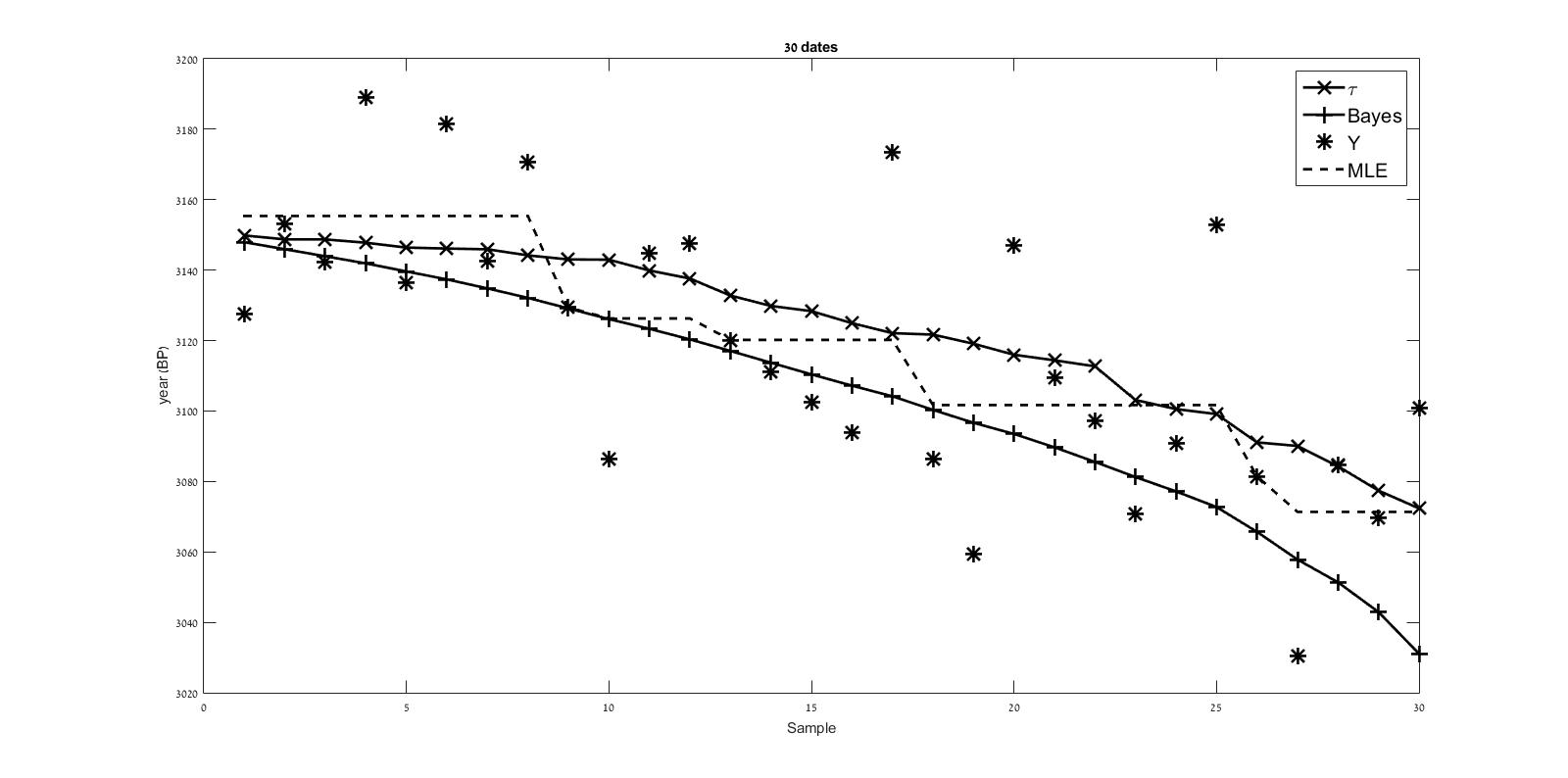}
\caption{Actual values, observations and Bayes estimator and MLE under order restrictions. Note that the Bayes
estimator is smoother than the MLE but considerably biased on the right hand side.}
\label{fig:BMLEmonotone}
\end{center}
\end{figure}

As an example, we simulated the following situation. 10 samples were taken, known to be in the interval 3150 BP
to 2850 BP. The actual values were a sequence of 10 years apart from 3140 BP to 3050 BP. The observations were
Gaussian with \sig =30. The Bayesian estimate are based on Monte Carlo Markov chain with K=1,000,000 steps. The
results are given in Table \ref{tab:1}.
\begin{table}[t]
\begin{center}
\caption{Bayesian analysis of a sequence of events}\label{tab:1}
\begin{tabular}{|d|d|d|c|}
\hline
\multicolumn{1}{|c|}{\mu (BP)}&	\multicolumn{1}{c|}{Y}&	\multicolumn{1}{c|}{Bayesian estimate}&	
\multicolumn{1}{c|}{Bayesian credible interval}\\
\hline
3140&	3130.2&	3137.2&	3149.5--3115.0\\
3130&	3116.3&	3144.2&	3144.3--3101.8\\
3120&	3050.0&	3113.4&	3134.7--3090.9\\
3110&	3136.8&	3106.8&	3128.2--3084.7\\
3100&	3080.3&	3098.5&	3120.0--3076.4\\
3090&	3088.4&	3091.3&	3112.9--3069.0\\
3080&	3076.3&	3084.3&	3106.3--3061.4\\
3070&	3111.7&	3077.6&	3100.3--3053.6\\
3060&	3088.8&	3068.7&	3093.1--3041.7\\
3050&	3129.2&	3057.3&	3085.1--3024.4\\
\hline
\end{tabular}
\end{center}
\end{table}
Clearly, the Bayesian estimate improves over the raw estimate of Y itself. The credible intervals cover the true
value in all cases. However, in a similar example, where we increase M to 30 and have the real times being
monotone but not evenly spaced,  a clear bias is introduced by the ``non-informative'' prior as is demonstrated
in Figure \ref{fig:BMLEmonotone}.

\subsection{Different layers}

We assume now that the observations came from $G$ consecutive layers. Formally, let $Y_{gm}$ be the $m$th observation from layer $g$, and assume as before that
$Y_{gm}$, $g=1,\dots,G$, $m=1,\dots,M_g$ are Gaussian independent random variables with a mean $\mu_{gm}$ and a standard
deviation $\sig_{gm}$ respectively. The scientific assumption that the layers are ordered is translated to the formal assumption that for any $k=1,\dots,M_g$ we have $\max_m⁡
\mu_{g-1,m}<\mu_{g,k}<\min_m⁡\mu_{g+1,m}$. There are many estimators and approaches that can be used to analyze
such data. The MLE is one of them. Another class of estimators is of Bayesian estimators. It is a class, as the
estimator depends heavily on the prior, even if it looks, prima facia, as a non-informative prior.

Suppose, for simplicity, that it is known that all of $\mu_{gm}$ are between
$t_s$ to $t_e$.  One simple minded prior assumes that  $\mu_{g,1},\dots,\mu_{g,M_g}$ are independent and uniformly
distributed between $\max_m⁡\mu_{g-1,m}$ to $\min_m⁡\mu_{g+1,m}$.

Another prior postulates  that the transition time between the periods are taken from the uniform distribution on
the interval between $t_s$ to $t_e$. Given the boundaries, say $\tau_g$ and $\tau_{g+1}$, the events $\mu_{g,k}$
are independent and have a given prior distribution on the interval $(\tau_g,\tau_{g+1})$, for example the uniform, or more generally, a scaled
beta.

\begin{figure}[t]
\begin{center}
\includegraphics[width=1\textwidth]{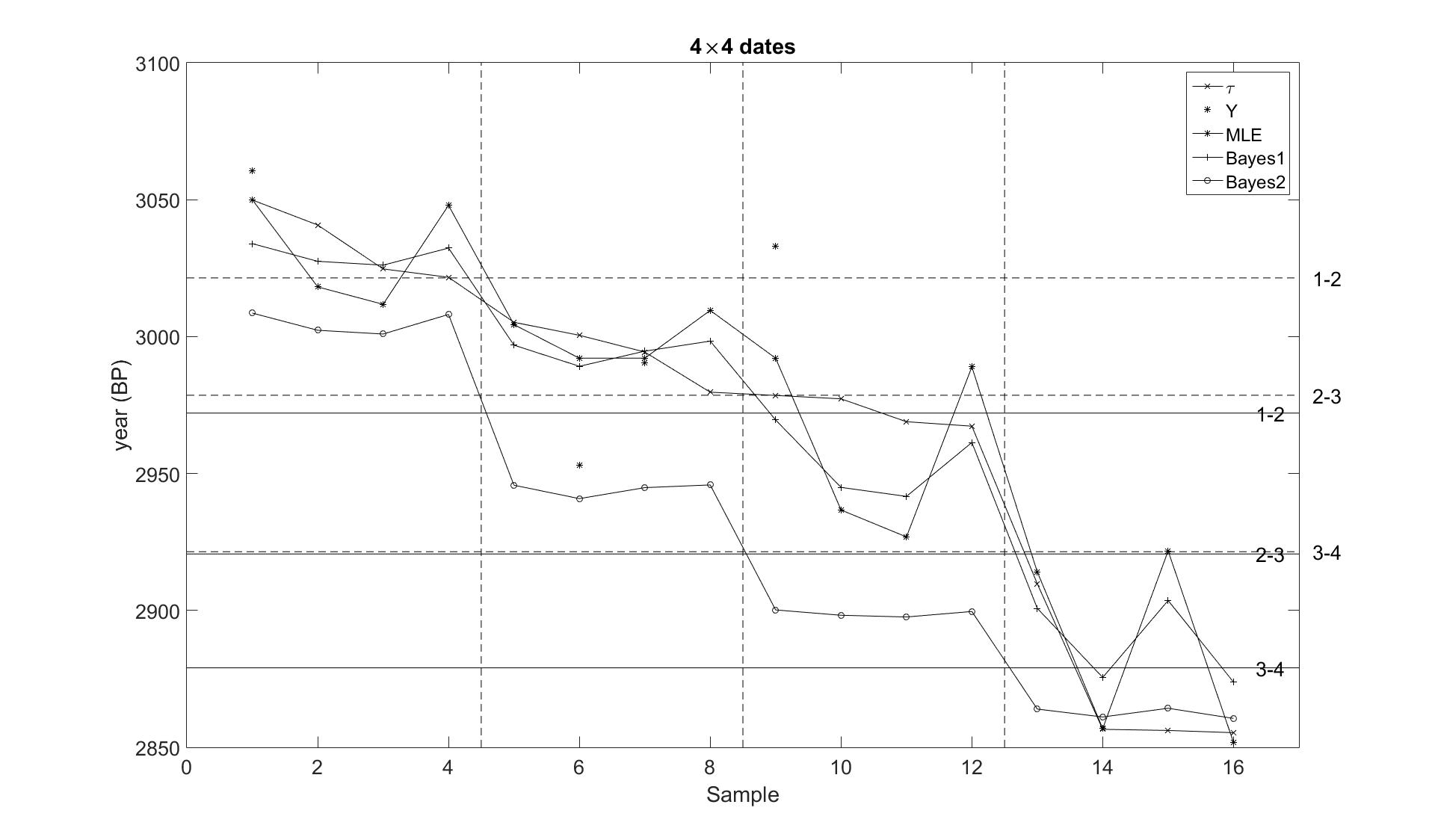}
\caption{4 laryer with 4 observations per layer. Data, MLE and two Bayes estimator. See text for details
A general model}
\label{4layers}
\end{center}
\end{figure}

These two priors seem to be quite similar, but in Figure \ref{4layers}  a typical example is presented. Four layers with four observations each  were sampled. This is a simulation, and therefore we know the ground
truth. The \mu's are known.  Random observations were drawn, and the three suggested above estimators were
calculated. The MLE is calculated using a modification of the polling adjacent violators algorithm, and the two
Bayes estimators were calculated using MCMC Green algorithm.
The horizontal lines denote the boundary points between the layers. The dashed lines are the true values while
the solid lines are those estimated using the second Bayes procedure. The vertical lines denote the division to
the four groups.  It can be observed  that the second Bayes procedure is strictly biased (all simulations
were essentially the same). The first Bayes estimator and the maximum likelihood estimator are very similar.

\section{A few words on the calibration curve}
\label{sec:calibration}
In our analysis so far we assumed that the the laboratory measures directly, albeit with error, the age of the finding, a seed, olive kernel or similar. However, this is not true. The laboratory presents its finding in an artificial date called \emph{Before Present} (BP). The translation from the BP to the standard calendar  date is done using the the calibration curve, which is based on \c14 dating of tree rings. One of the main difficulties with \c14 dating is that this calibration curve is non-linear, non-monotone, and measured with noise and smoothing (i.e., mostly on group of 10 years). For completeness, we discuss now a naively simple model for \c14 calibration.

The BP age is an artificial construct representing the measured level activity of \c14 (Stuiver and Polach 1977; Bronk Ramsey 2008). In fact, the BP age is just a complicated way to express the fraction of \c14
in the sample and is given by $\bp=-8033 \log ⁡f$, where $f$ is the fraction expected to be found in 1950 (the ``present'') relative
to the atmospheric concentration of \c14  in 1950. It was the age of the sample if three conditions were met: (1)
There were no measurement error; (2) The atmospheric concentration of \c14 at the  time of the sample was created was
the same as it was in 1950; and (3) The half-life time of \c14  was 5568 years (the Libby half-life).  No one of
these assumptions is correct. E.g., it is more likely that the half-life equals 5730 years (Godwin, 1962). The third problem is just a minor nuisance. The other two problems are real, not easy to correct and interplay.

\begin{figure}[t]
\begin{center}
\includegraphics[width=1\textwidth]{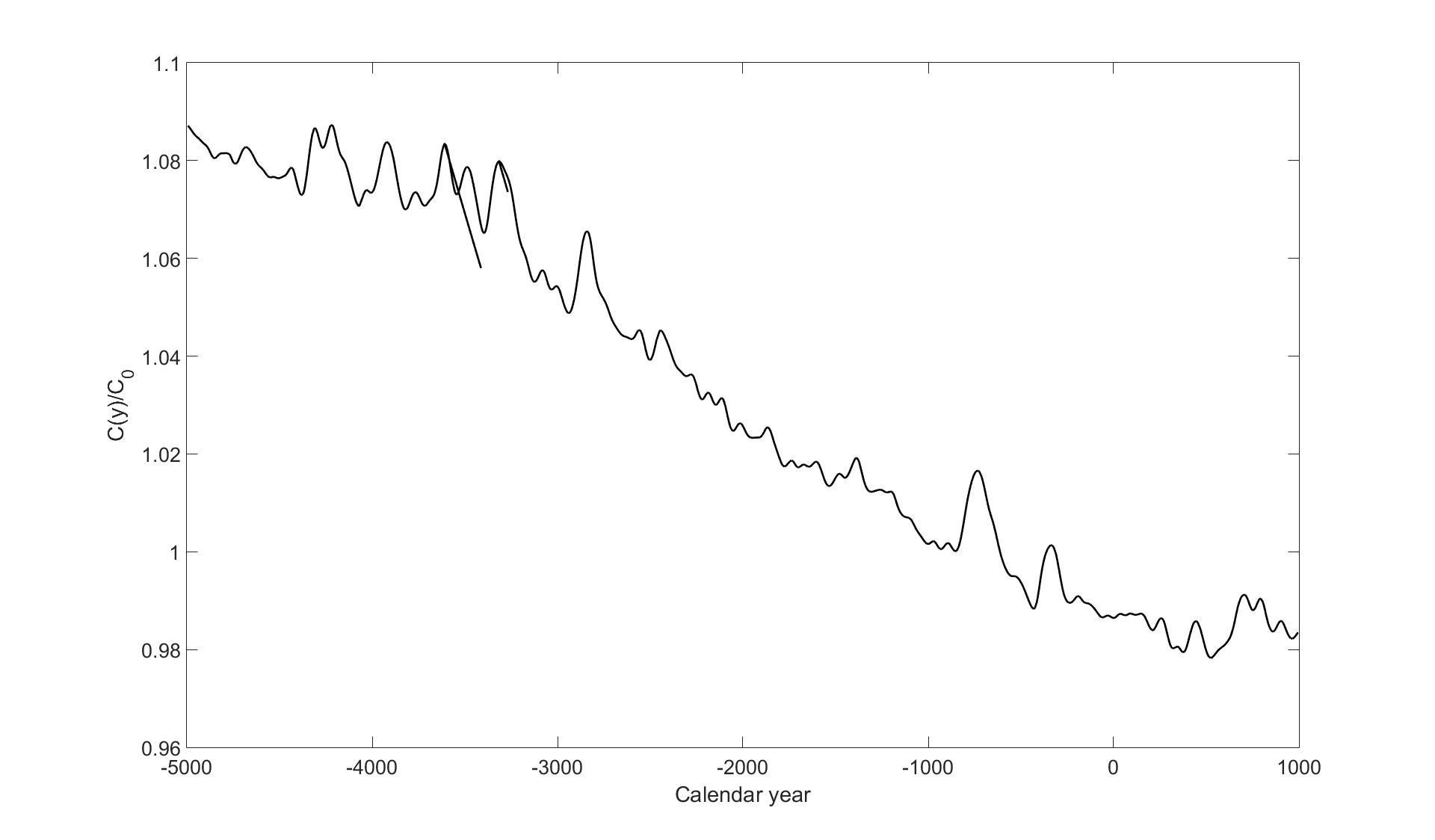}
\caption{Approximate concentration. IntCal13 data smoothed with a normal kernel with 20 years bandwidth
The plot is based on the data given by intCal13 (Reimer et al. 2013), data taken from
http://intcal.qub.ac.uk/intcal13/). The raw data points are smoothed using Gaussian kernel smoother with
bandwidth of 20 years. It can be observed that the curve start with  fluctuations  around a value somewhat higher
than 1, then within 3500 years or so drops to another stable value somewhat lower than 1.}
\label{fig:intCal13wide}
\end{center}
\end{figure}

 The concentration of \c14  in the atmosphere is not stable due to a few
processes  (Bronk Ramsey 2008). \c14  is generated in the upper atmosphere by nuclear reaction induced by
cosmic rays, masked by the Earth's magnetic field. The first is more or less constant, but the latter is subject
to temporal variation. The radiocarbon is removed from the atmosphere by the natural radioactive decay of the
isotope as well as injection of ``old'' carbon from water and underground. Again the latter is subject to
variation in time.  The result is a time varying concentration. The calibration data enable us to get an approximation of the history of \c14  reservoir. A good approximation is
given by
\eqsplit{
    C(y) e^{-(y-1950)/8267}=C_0 e^{-\bp/8033},
    }
or
 \eqsplit[c14conc]{
    C(y)/C_0 =e^{(y-1950)/8267-\bp/8033},
  }
where 8033 years is the Libby mean life and 8267 is the Cambridge mean life of \c14 , $C_0$ the fraction of \c14
at 1950, and $C(y)$ at year $y$. In Figure \ref{fig:intCal13wide} a plot of this
curve for six millennia is given.

\begin{figure}[t]
\begin{center}
\includegraphics[width=1\textwidth]{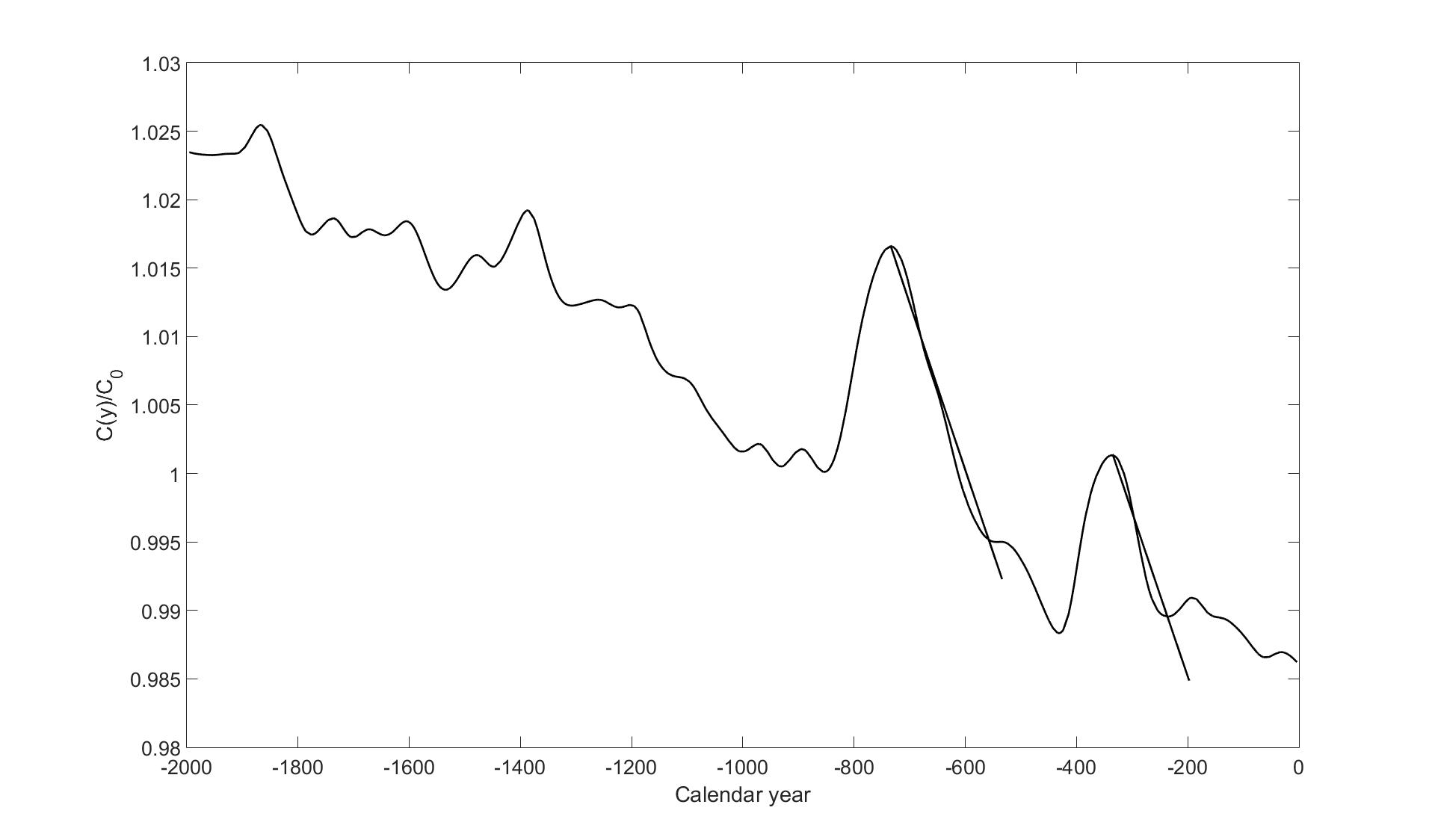}
\caption{Approximate concentration. IntCal13 data smoothed with a normal kernel with 20 years bandwidth. The
slops related to a pure \c14 decay are given for two points. }
\label{fig:intCal13Narrow}
\end{center}
\end{figure}

If we concentrate more on the relevant period to the Biblical times, we obtain Figure \ref{fig:intCal13Narrow}.
The period between 850 to 250 BCE is interesting. It has two periods of very high level of \c14  production, followed
by periods of fast drop. We added to the graph two lines showing what would be the reduction of the \c14
atmospheric concentration if  the only relevant active process after the peaks at 734 and 334 BCE would be the
radioactive decay. In particular, there would be no atmospheric generation of new \c14  atoms. As can be seen
from the figure, the actual decay is even faster, as if no new \c14  is generated and some old carbon is
injected.

Looking on equation \eqref{c14conc} differently, it gives the physical process that defines the calibration
curve:
 \eqsplit{
   \bp=\frac{8033}{8267} (y-1950)-\log\Bigl(\frac{C(y)}{C_0} \Bigr)\approx \frac{8033}{8267}
   (y-1950)-\frac{C(y)}{C_0} +.
  }
Thus the \c14  concentration process
is similar to the calibration process.

It may be reasonable to assume that the concentration is a result of many different independent processes,
neither of them dominates the rest, some of them were mentioned above, and different years would be independent.
It will be convenient to take as a prior for the calibration line the assumption that it was generated by a
Wiener process with a drift. It is a simple tool. Although the radioactive decay is
not constant but proportional to the current concentration (an Ornstein-Uhlenbeck process), but since the
half time life of \c14 is much longer than the time scales we consider in this paper, this effect can be ignored.

Does the $C(t)/C_0$  process really looks like a Brownian motion?  The increments do not seem as
having a stationary distribution, probably, the drift is not constant, and some fluctuations are larger than may be expected.

Nevertheless, posing  prior based on the assumption that $C(y)$ is a Brownian process with a drift seems to be
justified. Since the Brownian process is Markovian, the analysis of each century, say, is essentially independent and local, and the
lack of stationarity of the underline concentration is not really an obstacle.

But, is it really a robust assumption and does not bias the analysis as the Bayesians try to convince us? We test this using a simulated model of
reality. Mine conclusion from this simulation is that the Bayesian analysis should be considered with care. Here
are the details.   I will use the following notation and assumptions:

Let $y_1,y_2,\dots,y_n$ be the years for which we have calibration data, and $z_1,z_2,\dots,z_n$ be the
calibration data. Each observation is a Gaussian random variables with mean $\beta_i$ and variance $\nu_i$. The
vector $\beta_1,\beta_2,\dots,\beta_n$ is \emph{a priori}  assumed to be Gaussian, where the \emph{a priori} mean
of $\beta_i$ is assumed to be $\gamma_0+\gamma_1 y_i$, ($\gamma_0$ and $\gamma_1$ known for simplicity), and the
covariance of $\beta_i$ and $\beta_j$ is $\sig^2  \min⁡\{y_i,y_j\}$, $\sig^2$ unknown.  Thus
$\beta_1,\dots,\beta_n$ are a realization of a Wiener process with a drift.

 Gaussian Bayesian models are convenient because the \emph{a posteriori} distribution can be
easily calculated. Since both the statistical model and the prior are Gaussian, the \emph{a posteriori} distribution of
$\beta_1,\dots,\beta_n$ is Gaussian, and it is relatively simple to explicitly calculate it’s \emph{a posteriori}
mean --- the Bayesian calibration curve, and the \emph{a posteriori} variances and covariance of all pairs
$(\beta_i,\beta_j)$. However, Gaussian priors can be less naive and robust than one may assume, (Tuo and Wu 2016).

\begin{figure}[t]
\begin{center}
\includegraphics[width=1\textwidth]{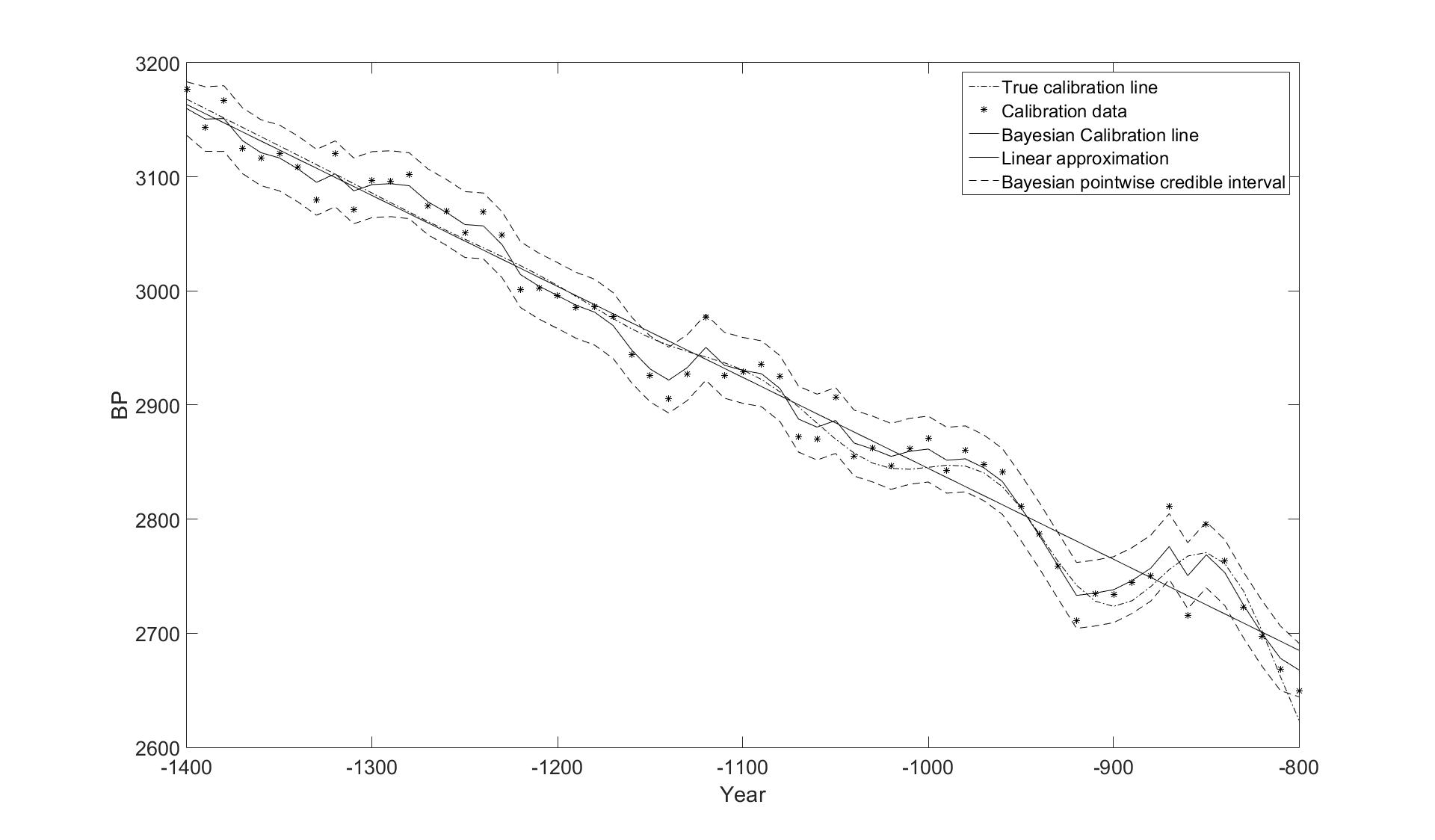}
\caption{Simulated data for calibration.}
\label{fig:simCal}
\end{center}
\end{figure}

I estimated the variance $ \sig^2$ using empirical Bayes concepts (Casella 1985) by looking for the value of
$\sig^2$ for which the sum of the \emph{a priori} second moments equals to its \emph{a posteriori} value.
In Figure \ref{fig:simCal} one simulation example is plotted, where the true calibration line was
\eqsplit{
    \beta_i=\gamma_0+\gamma_1 y_i+70 \Bigl(\frac{y_i-y_1}{y_n-y_1 }\Bigr)^3 \sin⁡\bigl(\frac{y_i}{20}\bigr) .
}
The ``measurements'' were taken every 10 years with a standard deviation of 21 years.

I then calculated estimates and 90\% confidence intervals for 5  `new observations' measured each with a
standard error of 20 years. For each observation the \emph{a posteriori} density was calculated and the credible
set was the set of years for which the density was above a threshold such that the total \emph{a posteriori}
probability was 0.9. The MLE was calculated and the frequentist confidence interval was the interval centered at
the MLE and with length equals to the total length of the Bayesian credible set. Note that the frequentist
confidence set is an interval, while typically the Bayesian credible set is not an interval but a union of a few
intervals, and hence its range may be larger than the frequentist one.  The estimates were simulated 5000 times
(with a new set of calibration data every 20 simulations). The results are given in Table \ref{tab:CIcc}. The
main conclusions from the table is that for the simulated model  the Bayesian credible sets are too large, and
the frequentist simpler model gives no worse coverage with shorter intervals.
\begin{table}
\begin{center}
\caption{Credible sets and confidence interval}
\label{tab:CIcc}
\vspace{1em}
\begin{tabular}{|r|dd|dd|}
\hline
&\multicolumn{2}{c|}{Bayesian credible set}&\multicolumn{2}{c|}{Frequentist confidence interval}\\
Year (BCE)&	 \multicolumn{1}{c}{coverage} & \multicolumn{1}{c|}{range (years)}&	 \multicolumn{1}{c}{coverage}
&\multicolumn{1}{c|}{range (years)}\\
\hline
1386 &  99.82\%  &   97.8  &  99.78\%  &   88.7 \\
1244 &  99.70\%  &  174.0  &  99.76\%  &  155.9 \\
1101 &  98.36\%  &  182.6  &  99.36\%  &  158.2 \\
 959 &  98.54\%  &  202.5  &  97.54\%  &  158.3 \\
 816 &  99.12\%  &  124.2  &  99.72\%  &   93.0 \\
  \hline
\end{tabular}
\end{center}
\end{table}

\begin{figure}[t]
\begin{center}
\includegraphics[width=1\textwidth]{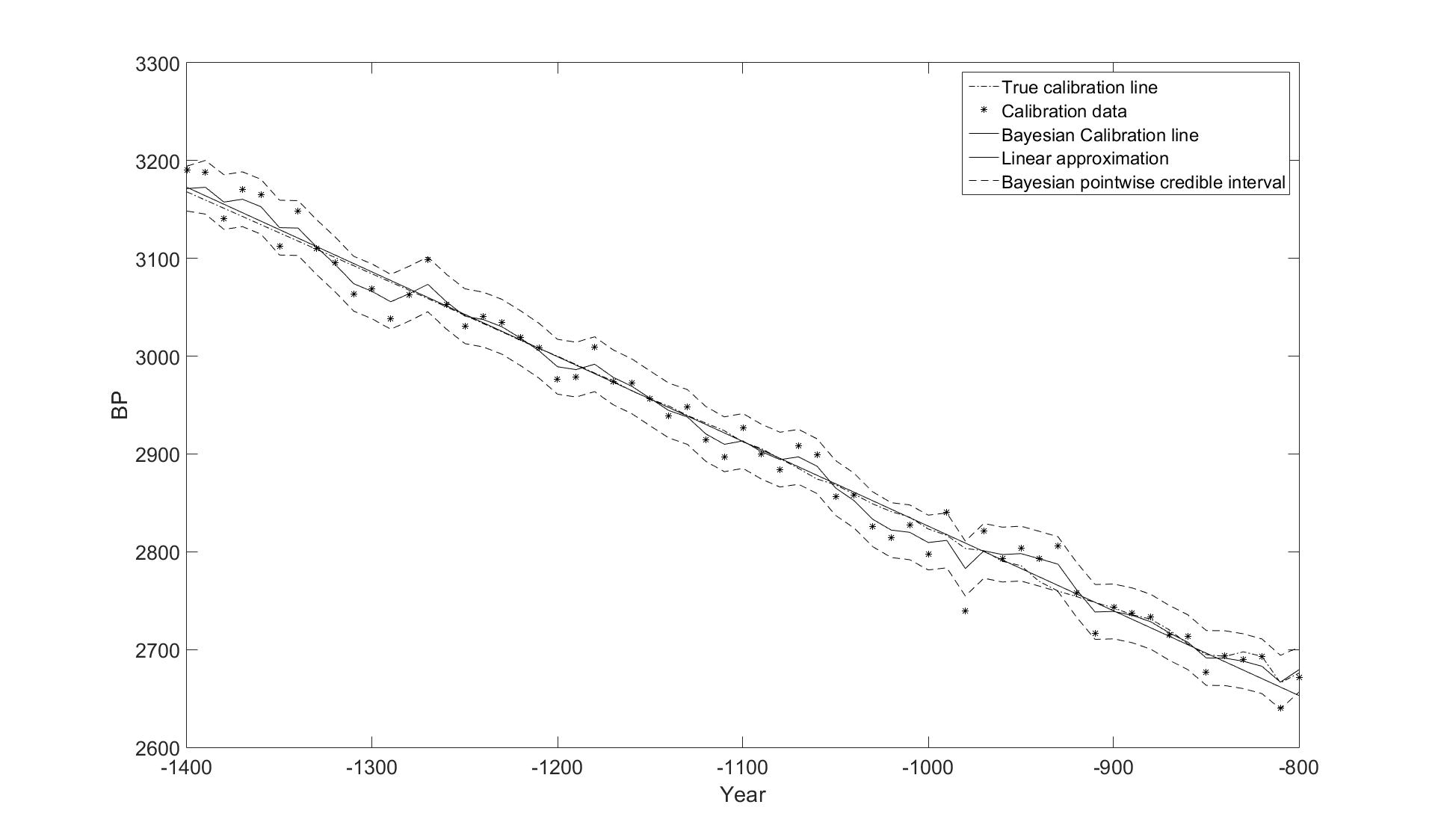}
\caption{Credible set and frequentist coverage for a few time points when the true calibration curve is samplled
from the prior.}
\label{fig:exPrCI}
\end{center}
\end{figure}

It could be argued that the over-coverage of the Bayesian credible set is because the `true' calibration curve
does not fit the prior. This is not the case. One problem with Bayesian credible sets is that their coverage is
as prescribed only `on the average', including the average over the possible states of Nature as prescribed by
the prior. In other words, the average is done over what the researcher has imagined to be possible.  Even in the
stationary situation we consider, their coverage depends on the real date of the sample.  Thus, their behavior is
expected to be different for samples taken at different times as the following example shows. The true calibration
curve was sampled now exactly from the prior. The Bayesian calculations were exact, but we obtained coverage
which depends on the time. Again we conducted a Monte Carlo experiment in which \emph{one} true calibration curve
was sampled from the prior, and 5000 observations on the calibration curve and `archaeological' samples were
taken.  See Figure \ref{fig:exPrCI} and Table \ref{tab:exPrCI} for the results. When I sampled from a proper
prior but with Gaussian process with variance changing in time (which fits the data better than stationary
variance), the results were even more extreme.

We conclude  from this analysis that the Bayesian analysis of the \emph{unique true} calibration curve is not justified,
and may lead to false credible sets. In Section \ref{ssec:RehovCal} the calibration curve for Tel
Rehov, restricted to the period surrounding the 10th century BCE is considered. I use there a simpler approach, which
is the one supported by the data and fits a standard good statistical practice.
\begin{table}
\begin{center}
\caption{Coverage probabilities of  random Gaussian true calibration curve }
\label{tab:exPrCI}
\vspace{1em}
\begin{tabular}{|r|dd|dd|}
\hline
&\multicolumn{2}{c|}{Bayesian credible set}&\multicolumn{2}{c|}{Frequentist confidence interval}\\
Year (BCE)&	 \multicolumn{1}{c}{coverage} & \multicolumn{1}{c|}{range (years)}&	 \multicolumn{1}{c}{coverage}
&\multicolumn{1}{c|}{range (years)}\\
\hline

1386 &  91.60\%  &   57.7  &  61.68\%  &   57.6 \\
1244 &  93.52\%  &   68.8  &  45.46\%  &   68.8 \\
1101 &  90.02\%  &  114.0  &  83.92\%  &  113.6 \\
 959 &  79.66\%  &  104.9  &  96.66\%  &  104.6 \\
 816 &  84.46\%  &   79.9  &  97.38\%  &   79.7 \\
 \hline
\end{tabular}
\end{center}
\end{table}

\section{Data analysis: Tel Rehov}
\label{sec:telRehov}

We move now to the analysis of the data from Tel Rehov
assuming that they represent 4 consecutive perriods, and the real interest is in the transition times. We start with a short introduction to the data we use.  Different researchers  faced the difficulty that the standard calibration lines of the tenth century BCE are not even monotone (Bruins et al. 2003; Mazar et al. 2005; van der Plicht and Bruins 2005; Sharon et al. 2007;  Mazar and Streit 2016). In the following we argue why we believe that by common statistical practice we can assume that this is not the case. Archeological data is susceptible to statistical gross errors, for example, seeds  that were found in an anachronistic layer. I discuss in Section \ref{ssec:robust}  a standard approach for dealing with outliers.    We then compare a frequentist and Bayesian analyses of the data.

\subsection{Backgraound}
\label{sec:rehovIntroduction}
The discussion of the absolute chronology of the Israeli Iron Age is of a special interest because of its  importance to the interpretation of biblical and extra-biblical sources  (Mazar 2005; Sharon et al. 2007).  The center of this debate is whether the Iron Age I/IIA transition happened early or late in the tenth century BCE. The importance of this dating stems from two assumptions. First, the description of the United Monarchy of David and Solomon could fit the  destruction of the tribal society of Iron I by David, and the construction  of the Iron IIA sites by Solomon.  Second,   the  inner biblical chronology dating with the existing extra biblical anchors, dates the Monarchy into the early tenth century, (Sharon et al. 2007).   The conflict is between the proponents of the high chronology that puts the transition at the early tenth century and makes the Biblical story feasible,  to those of believe in the low chronology that moves it up to the early ninth century contradicting the Bible (Mazar 2005; Sharon et al. 2007;  Finkelstein 1995, 1996). The difficulty with this periods is that areas like Greece and Cyprus lake  real chronological  anchors, and thus the chronology heavily dependeds on \c14 dating from the Israeli sites, (Boaretto et al. 2005).

A key anchor is the invasion by Pharaoh Shoshenq I
(Shishak) in 925 BCE. It is mentioned  both in
Egyptian inscriptions and the Hebrew Bible. The list of places raided by Shoshenq I,
mentioned at Karnak (Egypt), includes Rehov (Bruins et al. 2003). The Bible dates Shishak's invasion to 5 years after the death of Solomon.

Tel Rehov is the largest mound in the Beth-Shean Valley, 6 km west of the Jordan River, 5 km south of Tel Beth-Shean. It includes layers from the Late Bronze Age to the Early Islamic period.  Excavations of Tel Rehov were directed by Amihai Mazar of the Hebrew University of Jerusalem between the years 1997 and 2011,(Mazar 2013). Tel Rehov can be characterized by its dense stratigraphy of archeological levels from Iron-I and Iron-IIA, and by itself Tel Rehov provides the largest number of dates from any site in the Levant (Mazar and Streit 2016).  Our interest would be only with  the Iron I and IIA findings, and unlike Sharon et al. (2007), I do not consider it in the wider scope of findings from these periods found in other Israeli sites.

Several occupation phases were found from the Iron Age IB city (from the twelfth to the tenth centuries BCE), with different architectural characters. A massive building in stratum D-5 appears to be a storage building. Adjacent and later buildings in stratum D-4 were regular dwellings. Stratum D-3 is from the end of the period and includes more than 50 pits cut into the previous D-4 building and were probably used for food storage (Mazar  2013  and references therein).

Area C is characterized by successive floor layers of an open area. The findings include Iron-I pottery and pottery from the Coastal Plain (Mazar  2013  and references therein).

The main period exposed and studied is the Iron-IIA cty. Three general strata were, VI, V, and IV, were defined. A large number of radiocarbon dates of short-lived samples were measured, (Mazar  2013)  and references therein.

I used the data given in Mazar and Streit (2016) and described there in full. 161 determination were divided into  samples: R1 to R48. For some of them several repetitions were measured. The findings were analyzed in different laboratories (Rehovot, Tuscon and Groningen) and some of it by Iron Age Dating Project (Sharon et al. 2007).

\begin{figure}[t]
\begin{center}
\includegraphics[width=1\textwidth]{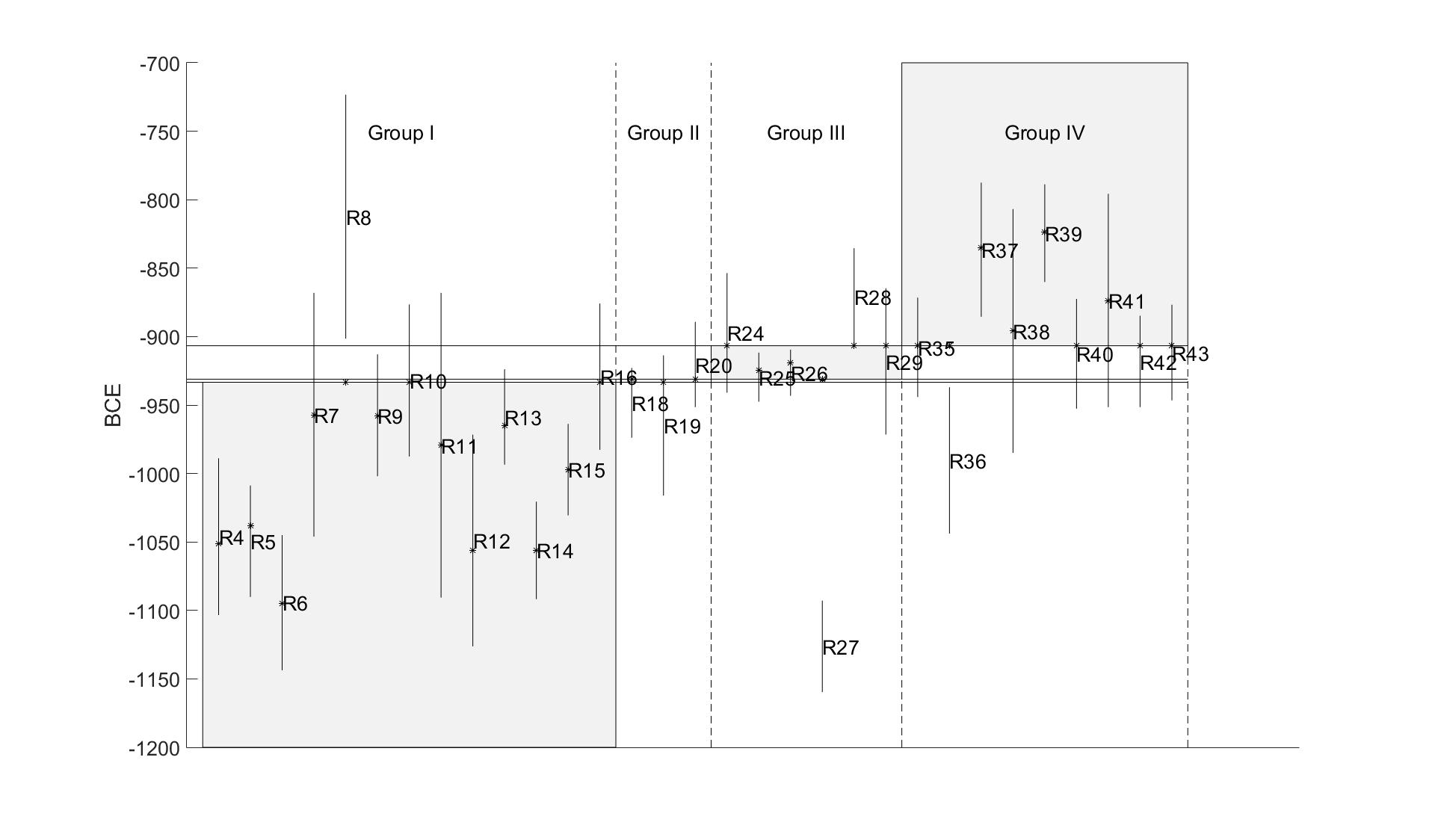}
\caption{Tel Rehov, data and MLE (see Section \ref{sec:mle} for details).}
\label{fig:TelRehovData1}
\end{center}
\end{figure}

The data set I analyzed is composed from 32 samples, and a total of 86
measurements.
The data are grouped into 4 groups, assumed to be consecutive, see (Mazar 2005; Bruins et al. 2005; Mazar 2013; Mazar and Streit 2016).\footnote{Omitted from this
discussion are Samples R1-R3 from Stratum D-6,
Sample R17, an outlier from Area B,  Samples R21 and R22 which are either from D-3 or D-2, R23 which is either from D-2 or D-1, and R30-R34 from Building CG, where attribution to either
Stratum C-1b or C-1a was not decisive.}:
\begin{itemize}
  \item
	Stratum D-4 and D-3 (R4--R16). Area D includes layers from the Late Bronze Age to the Iron Age IIA. In Stratum D-4 a well defined Iron Age IB pottery was found. Floor surfaces were excavated in the street as well as a court yard. The floor surface was raised from time to time.
\item
	Stratum C-2 (or stratum VI, R18--R20). In Area C building destroyed with their content were found, including two layers of destruction layers. The three strata, VI--IV includes mudbricks building. Stratum VI includes at least four main units.
\item
	Stratum C-1b (stratum V, R24--R29). The stratum includes an apiary and its surroundings as well as several other structures. The buildings were build with wooden beams, and it was destroyed by a fierce fire.
\item
	 Strata C-1a, B-5 and E-1 (strata V or IV, R35--R43). This period ended by another catastrophic event and abandonment. The finding includes characteristic pottery.
 \end{itemize}
The full set of observations is also presented in Figure \ref{fig:TelRehovData1}. Each sample is presented as a line centered at the mean
of the measurements, and length which is 4 standard errors of the mean as reported by the laboratories (i.e.,
approximately the 95\% confidence interval of the laboratory measurement).

\subsection{The calibration line for the tenth century BCE}
\label{ssec:RehovCal}
\begin{figure}[t]
\begin{center}
\includegraphics[width=0.45\textwidth]{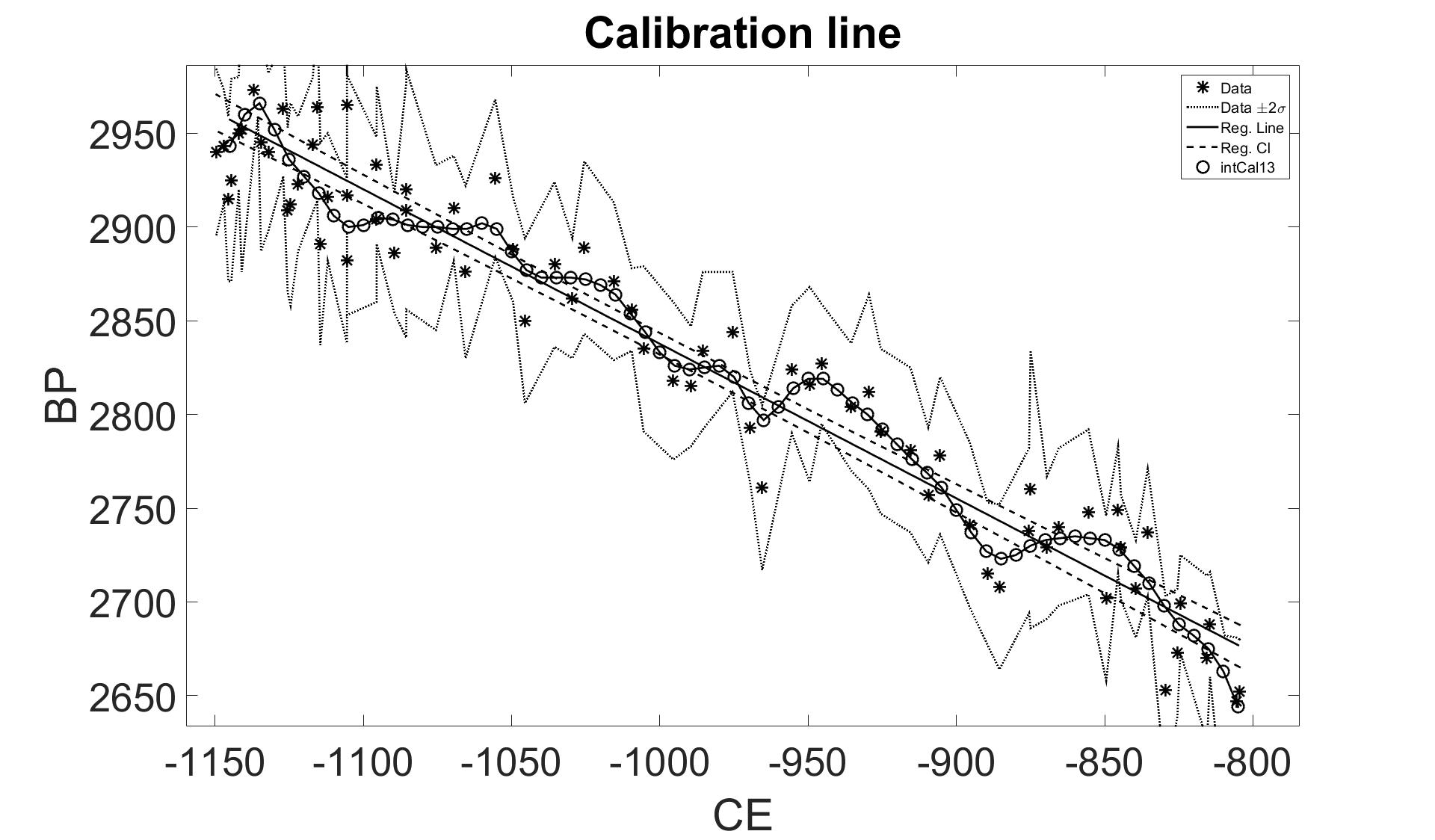}
\includegraphics[width=0.45\textwidth]{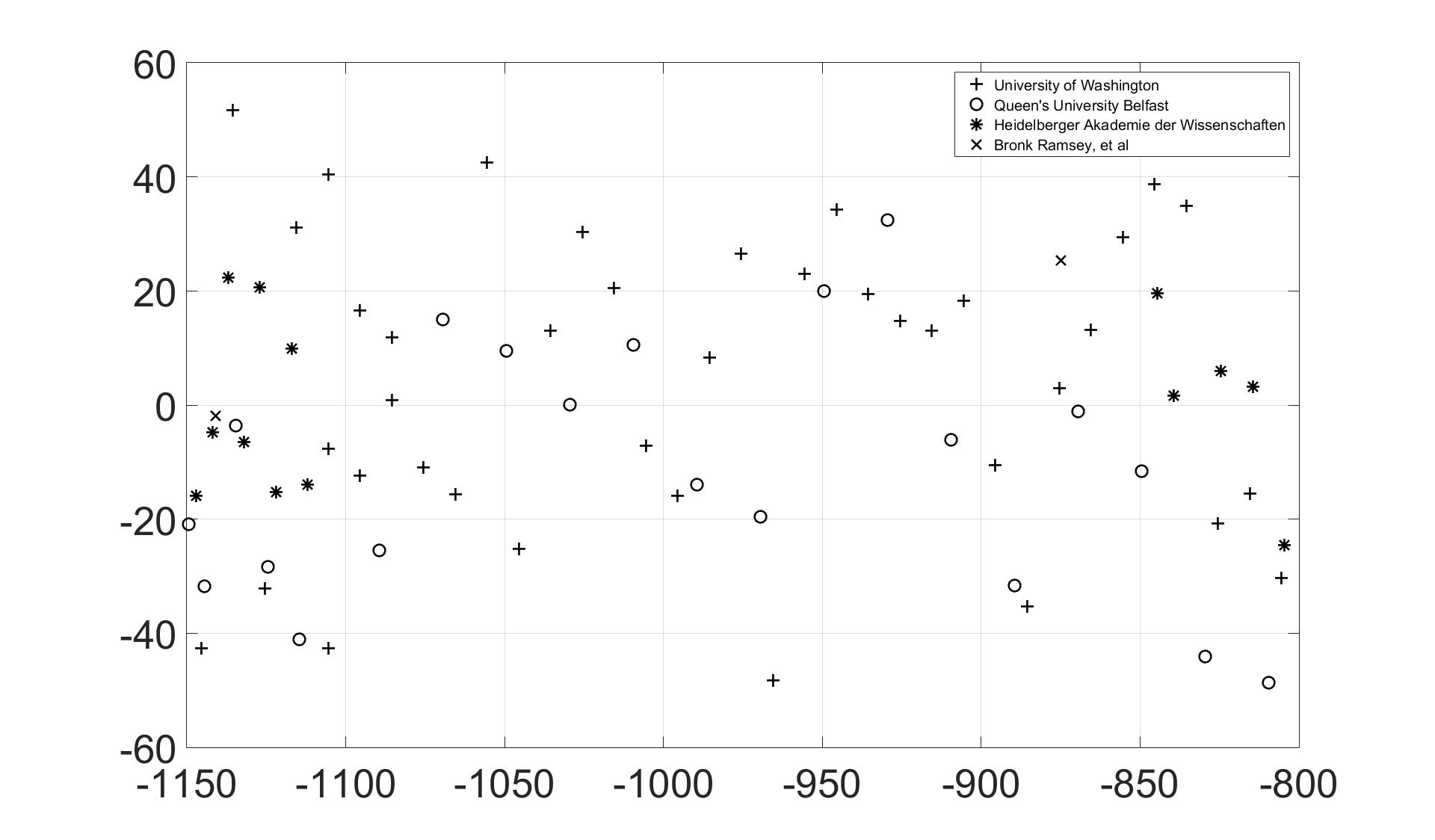}
\caption{The calibration curve for the Iron I--Iron IIA period. Left: The calibration data (starts, intCal13 all
data sets, http://intcal.qub.ac.uk/intcal13/) with 2 sigma confidence bound, the linear approximation, and
intCal13 calibration line (circles) for this period. Right: the residuals from the linear regression line,
markers by lab.}
\label{calCurve10th}
\end{center}
\end{figure}

The raw dates in Figure \ref{fig:TelRehovData1} are expressed in the Gregorian calendar
years. Unlike the common practice I use a simple calibration formula $Y_{ce}=2221.8-1.135*Y_{bp}$, where
$Y_{bp}$ is the laboratory measurement in the standard `before present' units, and $Y_{ce}$ is the year in the
Gregorian calendar. This scaling is based on the linear regression approximation of the calibration graph around
the time analyzed. I argue   that this is the common statistical practice, which follows  Occam's razor, since
the data do not support a more complex calibration curve. The argument is based on Figure \ref{calCurve10th}. In
these graphs  I present the regression of the data points of intCal13 (Reimer et al. 2013), in the relevant
interval between -1150 to -800 BCE excluding two isolated outliers at 997 and 1109.5 BCE (which will be tuned down by any robust analysis, see Section \ref{ssec:robust}.  In particular, I would
mention that the   serial correlation of the residuals, 0.14, is not significantly different from 0  (P-value of 0.11, one sided
permutation test), and the P-value is 0.25 if we
permute separately the two chronical halves of the data. If anything, there is a significant lab effect (of approximately 14 years).  Thus the data do not reject the linear model with independent error in favor
of the more general random walk model of the type considered in intCal13 (Blackwell and Buck 2008; Heaton et al. 2009; Niu et al. 2013;  Reimer et al. 2013).   Moreover, it is enough for
the validity of the argument that the real relevant calibration curve (if there is one) is just  monotone. We should, however, emphsis: some authors worry that the curve is not mononote (Bruins et al. 2003; Mazar et al. 2005; van der Plicht and Bruins 2005; Sharon et al. 2007;  Mazar and Streit 2016). For the
methodological discussion of this paper, it is certainly enough that it simplifies the discussion without any
serious impact on the claims made.

.
\subsection{Robust analysis of the data}
\label{ssec:robust}
It seems obvious that there are outliers in the data. For example, R8 seems to be too new for the stratum, R27 and R36 are too
old.  Thus, R27 seems to be approximately 200 years or 13 standard errors before its period.  Therefore,
assuming a simple minded Gaussian model may be dangerous --- a Gaussian model  puts too heavy weights on remote
outliers.

In line with the common statistical practice, I avoid the rejection of outliers, but do not ignore their presence (Huber 2011).
The general estimator I consider in the following replaces the mean by the solution $\hat\mu$ of
$\sum_{i=1}^m\psi_{c}(Y_i-\hat\mu )=0$, assuming all standard errors are the same (see below for the
generalization I use). If $\psi_{c}(x)\equiv x$, then $\hat\mu$ is the mean. If $\psi_{c}(x)=1$ for $x\ge0$ and
$-1$ for $x<0$, then $\hat\mu$ is the median. More generally $\psi_{c}(x)=x$ for $|x|\le c$, $\psi_{c}(x)=c$ if
$x>c$ and $-c$ if $x<-c$.
This estimator is the MLE if we assume that the density of the observations follows a density which is like a
Gaussian in the center but with somewhat heavier exponential tails.  
Again, the extreme cases are the normal ($c=\en$) and the double exponential ($c=0$).

The rationale of using an estimator based on such a generalization is based on the boundness of $\psi_{c}$ --- a
large outlier has a bounded effect on the estimator. Huber (1964) proves that $\psi_{c}$ gives an optimal
protection against an adversary who can place a few of the observations everywhere he wants, but in a symmetric
way around the true value. The effect of using $\psi_{c}$ is that remote points are not removed but pulled closer
to the center.
\subsection{The MLE}
\label{sec:mle}
The maximum likelihood estimation proceeds in two steps. In the outer loop, it places the boundary between the
periods, and in the inner loop,  it maximizes over the time value of the each sample given the boundaries of its period. Suppose the sample is composed
of $Y_{gm1}$, $g=1,\dots,G$, $m=1,\dots,M_g$, $i=1,\dots,I_{gm}$ of laboratory measurements, with standard errors
$\sig_{gm1},\dots,\sig_{gmI_{gm}} $ respectively, where $g$ denote the period (stratum), $m$ the sample, and $i$
the measurements.  Let period $g$ be with boundaries $\tau_g<\tau_{g+1}$.  If the (unique solution) of
$$A_{gm} (\mu)=\sum_{i=1}^{I_{gm}}\frac{1}{\sig_{gmi}}\psi_{c}(\frac{Y_{gmi}-\mu}{\sig_{gmi}} )=0$$
is between $\tau_g$ to $\tau_{g+1}$ then this is the estimate $\mu_{gm}$. If $A_{gm} (\tau_{g+1} )>0$ then the
estimator is $\tau_{g+1}$, and finally if $A_{gm} (\tau_g )<0$ then the estimator is $\tau_g$. Since $A_{gm}
(\mu)$ is a decreasing function of \mu only one of these can happen, and hence the estimator is well defined.
Denote the estimator by $\hat\mu_{gm}({\boldsymbol \tau})$.

Finding the estimators $\hat\mu_{gm}({\boldsymbol \tau})$, $m=1,\dots,M_g$, where $M_g$ is the number of samples
and ${\boldsymbol \tau}=(\tau_1,\dots,\tau_G)$, we can proceed to the next stage. First the (profile
pseudo) log-likelihood of the samples is calculated:
\eqsplit{
   l(\tau_2,\dots,\tau_G) &= \sum_{g=1}^G \sum_{m=1}^{M_g}
   \sum_{i=1}^{I_{gm}}\rho_{c}(\frac{Y_{gmi}-\hat\mu_{gm}({\boldsymbol \tau})}{\sig_{gm}} ),
  }
where $\rho_{c}(x)=-\log \bigl(f_c(c)\bigr)$ is the function whose derivative is $\psi_{c}$.  Then  the
values of $\tau_2,\dots,\tau_G$ that maximize l are found.

The maximizing values of $\tau_1,\dots,\tau_4$ are given by the horizontal lines of Figure \ref{fig:TelRehovData1}. The vertical broken lines denotes the 4 groups of observations. Thus, if the model is
correct, the observations were supposed to be all in the gray areas. The stars denote the values of $\hat\mu_s$.
Thus, if a star is on the boundary, the observation is truncated. If the full line of the observations is outside
the gray area, this is an outlier.

\begin{figure}[t]
\begin{center}
\includegraphics[width=1\textwidth]{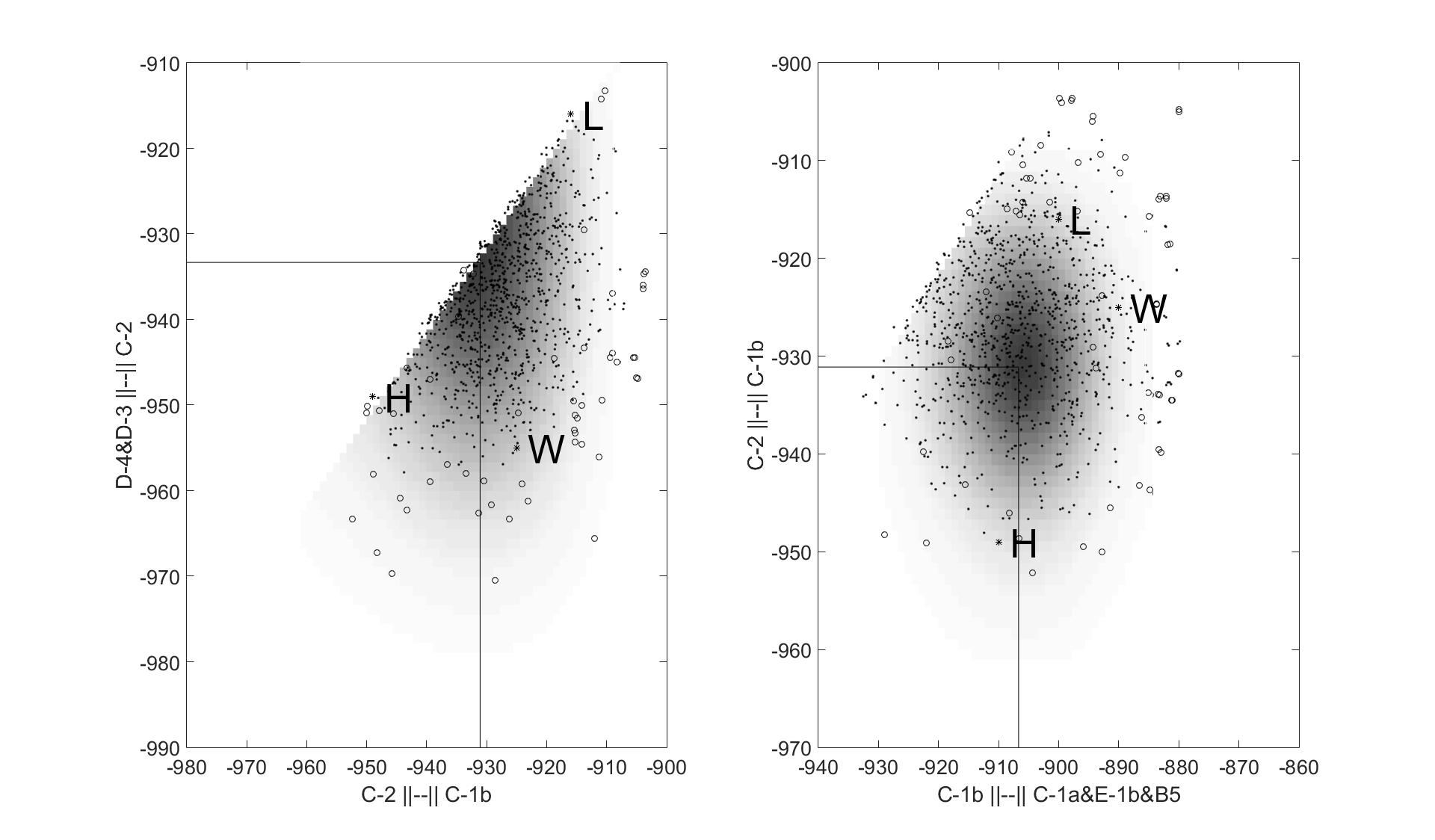}
\caption{Profile likelihood curve of the the two pairs of boundaries. The darker the area, the likelihood is
higher.  Left: the profile likelihood as function of the transition time between D-4 \& D-3 to C-2 and the
transition time between C-2 to C-1b.   Right: the transition between C-2 to C-1b and the transition time between
C-1b to C-1a, B and E. The marked stars denote the parameters of the detailed solution given in Figure
\ref{fig:highLowWide}. The dots and the circles are the maximizers of the bootstrap evaluations. See text for
details}
\label{fig:profLike}
\end{center}
\end{figure}

The profile likelihood of each pair of boundaries is given in Figure \ref{fig:profLike}. These graphs describe
the likelihood surface as function of the time of the two boundaries, after maximizing over all the other
parameters (the other boundary and the sample times).  The darker the color, the higher the likelihood. The
vertical and horizontal lines denote the profile maximum likelihood estimators.

\subsection{Confidence sets and the bootstrap}

It may seem that Figure \ref{fig:profLike} indicates that the transition between D-4 and D-3 to C-2 was around
932 BCE, the transition from C-2 to C-1b was soon after, and then the transition to E-1b and B-5 was around
906 BCE. However, these are just the maximizing values of of the likelihood function. There is, of course, a lot
of noise and uncertainty in these estimators, and they depend too much  on the particular random values measured.
To gauge how much uncertainty there is, I conducted  bootstrap simulation studies.   Aczel (1995) suggested the application of the bootstrap to archeological data.

The logic of the bootstrap is that the error distribution is a smooth function of the true data distribution,
hence one can evaluate the error in his inference by considering the error under distribution that is close to
the data distribution. There are two main classes of the bootstrap used in practice and  below. The nonparametric
bootstrap in which one samples a new sample by sampling with replacement from the observed sample or some
variation of this, and the parametric bootstrap in which one samples from an estimated parametric model. The
bootstrap analysis is usually done under the assumption that the model is regular. However, we deal in this
analysis with an irregular model, in particular the maximum of the likelihood is on the boundary of the parameter
set (because of the order constraints).  Yet, one can consider the bootstrap as a bagging procedure (Breiman
1996). One does not want that the scientific conclusions would depend on the existence or nonexistence of a
particular finding, one out of the 31 found. The conclusion should be robust to small changes in the composition
of the random items unearthed.

Under the nonparametric bootstrap 5000 random sampled were drawn. Each time, I kept the number of sub-samples
fixed at the value of 31. However I sampled without replacement from the 31 found sub-samples. If one of strata
was empty at this step, which happened with probability of 0.04, I randomly added one of its sub-samples.  Thus,
on the average, each sub-sample appeared slightly more than once, but it could appear twice or none at all. For
each bootstrap sample the maximum likelihood was calculated. The next step was locating the 4750 more likely of
them. This was done by considering the 4750 points that fall inside the ellipsoid based on the principle
components analysis (PCA) of the correlation matrix of the 5000 estimators (see below) and includes 95\% of them.
Similar results were obtained by other nonparametric bootstrap schemes (e.g., Poisson sampling of sub-samples, or
keeping the strata sizes fixed).

The nonparametric bootstrap is justified for simple random samples, which is not the case here. To see how it
could go wrong, suppose we had only two periods. Suppose there are 30 sub-samples from the middle of the eleventh
century BCE belonging to the first period, 30 sub-samples from the middle of the ninth belonging to the second
period, and one from each period dated by the lab to the middle of the tenth century. In this case, the two
sub-samples from the tenth century  are the only ones that are informative, but from 60\% of the bootstrap samples at least one of them will be
missed. Thus, the bootstrap would give a misleading picture. This extreme
situation is very different from that of Tel Rehov, but to be safe I used also the parametric scheme. Here we
sampled observations from the normal distributions with standard distribution as given in the data and means as
estimated by the MLE. Again, using PCA of the estimates, the central ellipsoid with 950 bootstrap values was
found.  The  points within this ellipsoid are marked on Figure \ref{fig:profLike} as dots, while the circles are
the remaining 50 points.  Note that the ellipsoid is 3 dimensional, and hence in the two dimensional projections
of the figure there seems to be some mix of the dots and circles.

\begin{figure}[t]
\begin{center}
\includegraphics[width=1\textwidth]{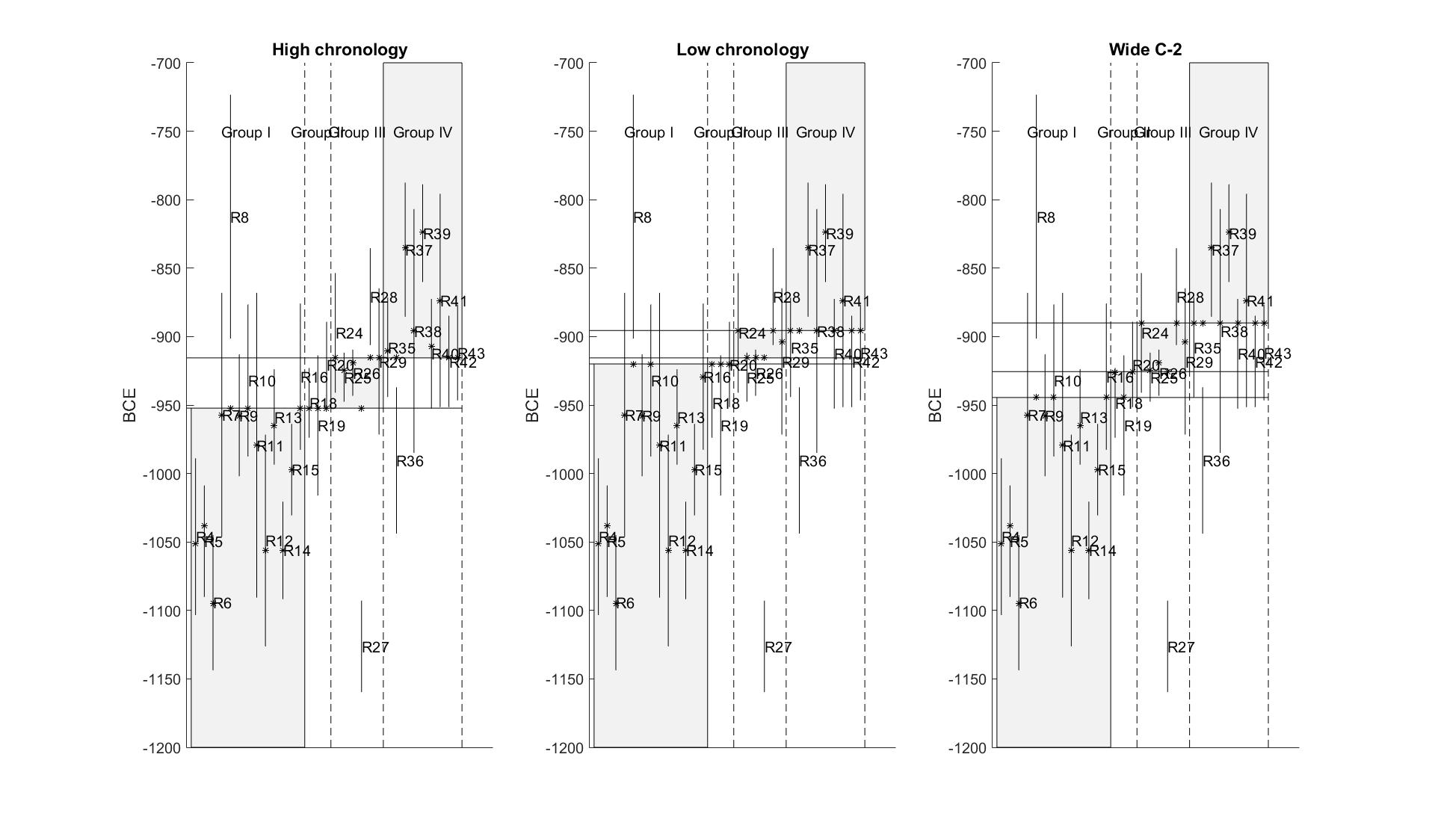}
\caption{Possible chronologies for Tel Rehov, evaluated at the points marked in Figure \ref{fig:profLike}.}
\label{fig:highLowWide}
\end{center}
\end{figure}

The meaning of this cloud of points is that any point within the ellipsoid gives a possible scenario, and the
available data cannot differentiate between these. Three points were marked in Figure \ref{fig:profLike}, which
correspond to a high chronology, low chronology, and a chronology with a relative long C-2 period. These 3 points
are well inside the above mentioned ellipsoid. The chronologies themselves are given in detail in Figure
\ref{fig:highLowWide}.  The data seem to support these chronologies, and only scientific (i.e., archaeological)
information external to these data can be used to decide between them.

\begin{figure}[t]
\begin{center}
\includegraphics[width=1\textwidth]{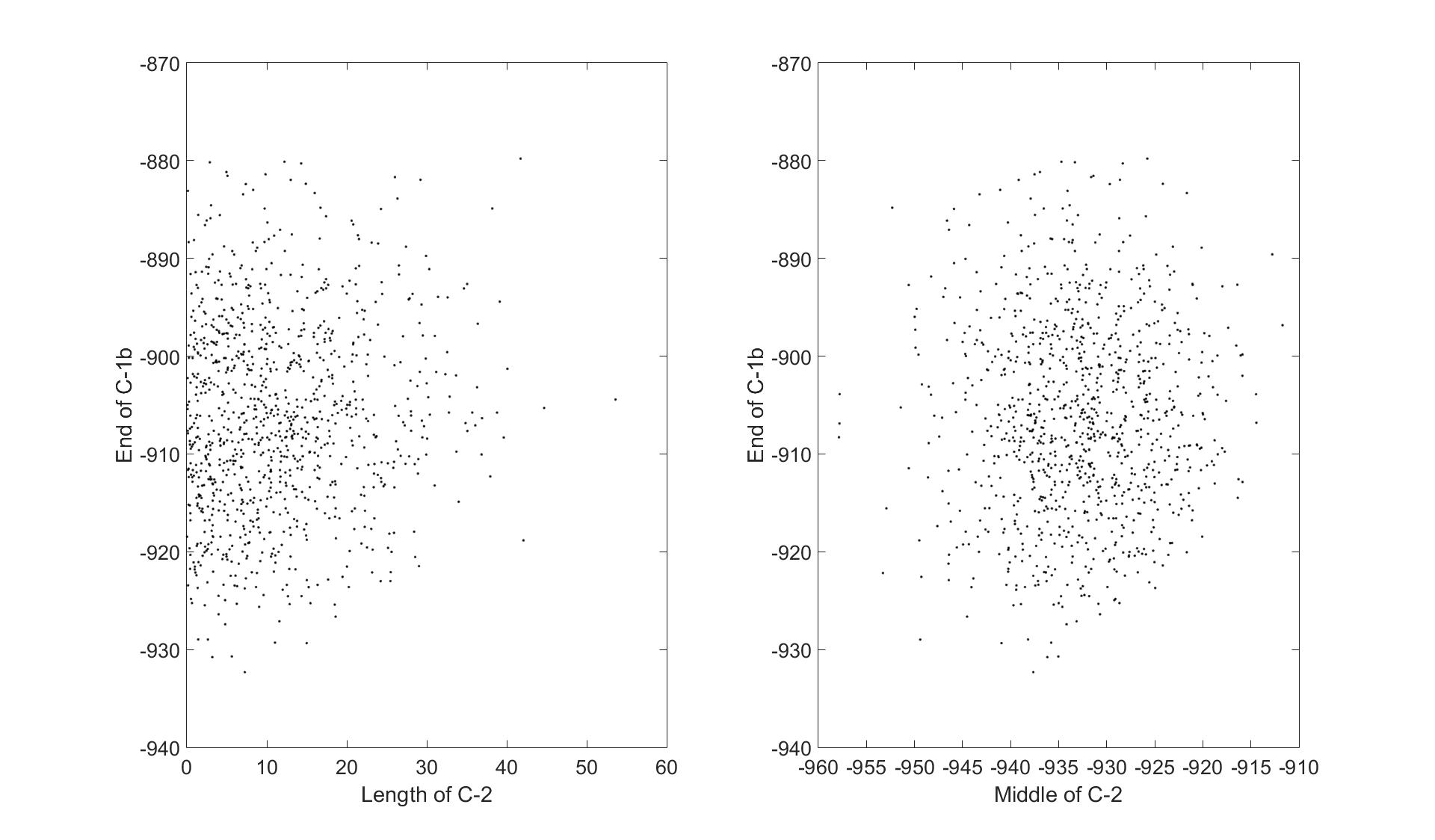}
\caption{Principle component analysis of the parametric bootstrap estimators, first 1000 observations.}
\label{fig:PCA}
\end{center}
\end{figure}

It is interesting to consider the principle component analysis (PCA) of the bootstrap estimate.  The 3 new variables found
based on the PCA and intervals that include 95\% of their values were, essentially: (1) the length of the C-2
stratum (0 to 21.5 years). (2) The end point of C-1b (920 to 893 BCE), and, (3) the middle time of the C-2
stratum (950 to 922 BCE.).\footnote{The exact normalized eigenvectors  of the nonparametric bootstrap correlation
matrix were:  $(0.71,-0.71,  0.02)$, $( 0.03, 0.01,-1.00)$, and $(    0.71,   0.71,  0.03)$, while those of  the
parametric bootstrap correlation matrix were: $( -0.70,    0.70,   -0.15)$,
$( -0.15,    0.06,    0.99)$, and $( -0.70,   -0.71,   -0.07)$.}  The first 1000 bootstrap realizations of these
variables are given in Figure \ref{fig:PCA}. One main conclusion from this analysis is that the start and end of
C-1b can be analyzed independently.

\subsection{Bayesian analysis of Tel Rehov data.}
Bayesian analysis of these Tel Rehov data is simple. Its main strength is that it solves elegantly the problem of
confidence sets. Credible sets for any parameter or group of parameters are automatic, intuitive
and efficient on their own terms.

\begin{figure}[t]
\begin{center}
\includegraphics[width=1\textwidth]{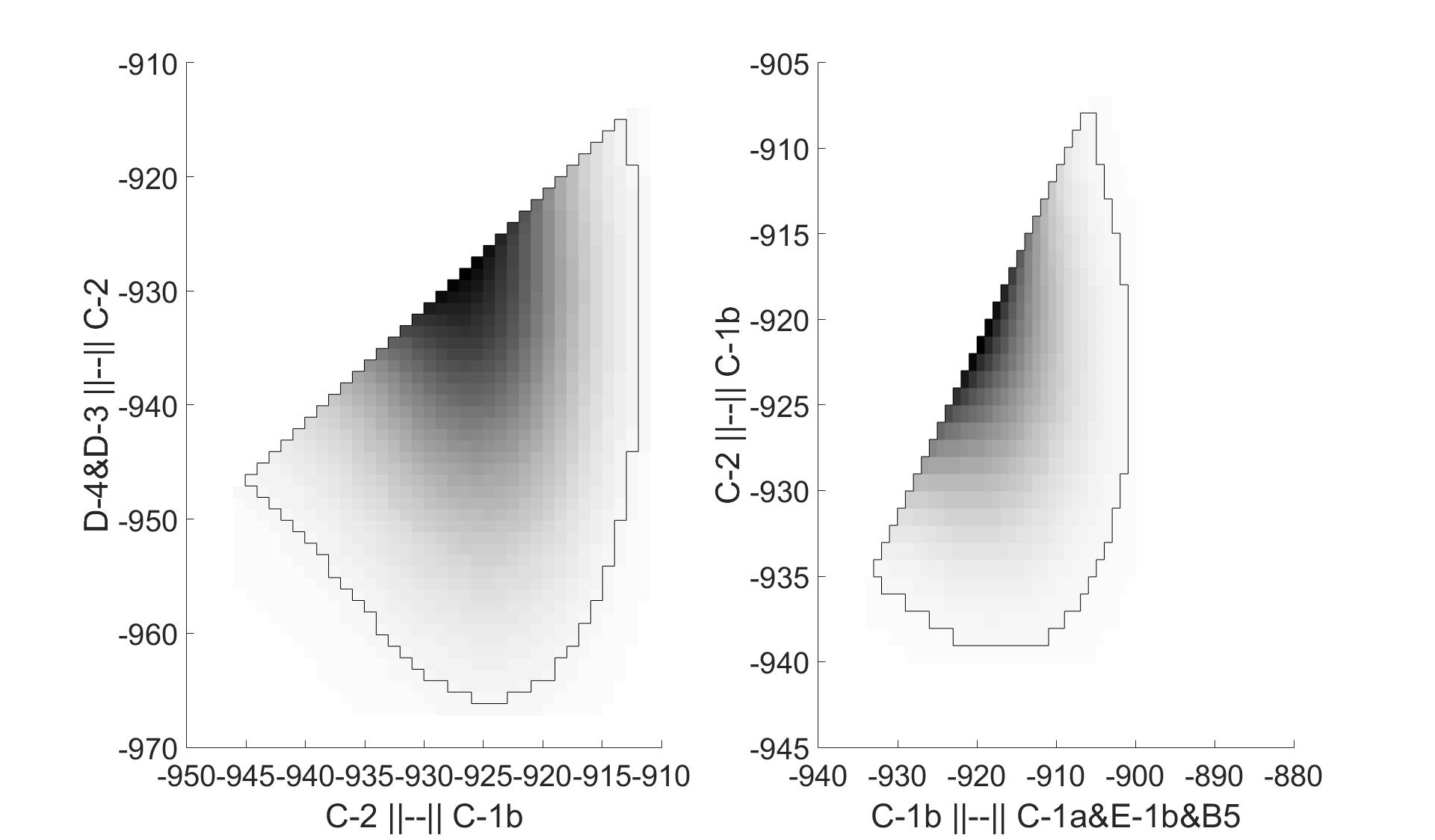}
\caption{Bayesian \emph{a posteriori} density of the two pairs of transition times. The darker color the higher
the density. Black line is the 95\% credible set. Left: the joint \emph{a posteriori} density of the transition
time between D-4 \& D-3 to C-2 and the transition time between C-2 to C-1b.   Right: the joint \emph{a
posteriori} of the  transition between C-2 to C-1b and the transition time between C-1b to C-1a, B and E.}
\label{fig:bayesRehov}
\end{center}
\end{figure}

Our prior  is built in two stages. The first is the \emph{a priori} assumption on the transition times. I take
them to be uniformly distributed on their domain of definition (the cone of ordered three dimensional vectors).
Given the transition times, I assume, as is quite natural, that the times of the samples are independent uniform
on the interval of their definition. Thus the \emph{a posteriori} distribution of the 3 boundaries is
 \eqsplit[pss]{
    \pi(\tau_1,\tau_2,\tau_3|data)=c\prod_{s=1}^4 \prod_{m=1}^{M_s} \Bigl( \frac{1}{\tau_s-\tau_{s+1}}
    \int_{\tau_{s-1}}^{\tau_s} \prod_{i=1}^{I_{sm}} e^{\rho_{c}(\frac{Y_{smi}-t}{\sig_{sm}})}dt\Bigr)
  }
The normalizing constant is used to ensure that $\pi(\tau_1,\tau_2,\tau_3|data)$ integrates to 1.
The results of the Tel Rehov data are presented  in Figure \ref{fig:bayesRehov}.

The black boundaries in this figure denote the smallest sets with 0.95 a posterriori probability (i.e., they are
sets defined by $$S_g (p)=\{\tau_g,\tau_{g+1}: \pi_{post} (\tau_g,\tau_{g+1} )>p\},$$ where $p$ is the solution
of $\dint_{S_g (p) } \pi_{post} (s,t)ds dt=0.95$.

If we concentrate on the boundary between C-2 to C-1b, then its value in the bootstrap values within the 95\%
ellipsoid are in the range of 948 to 907 BCE, see Figure  \ref{fig:profLike}. The credible set is
similar,although somewhat earlier and longer. The projection of the credible set in Figure \ref{fig:bayesRehov}
is  966 to 915 BCE. Thus, both analyses indicate that the transition is  likely to be at the first part of the
second half of the tenth century BCE, but it may that it was somewhat earlier. However, the right panels in
figures  \ref{fig:profLike} and \ref{fig:bayesRehov} are very different. The  transitions between C-2 to C-1b and
C-1b to C-1a\&E-1b\&B5 seem to be almost independent by Figure \ref{fig:profLike}, while they are far from that
in Figure \ref{fig:bayesRehov}.    In fact, by the Bayesian analysis C-1b is likely to be very short --- the
darkest area in the right panel of Figure \ref{fig:bayesRehov} is close to the boundary of zero length C-1b.

\begin{figure}[t]
\begin{center}
\includegraphics[width=1\textwidth]{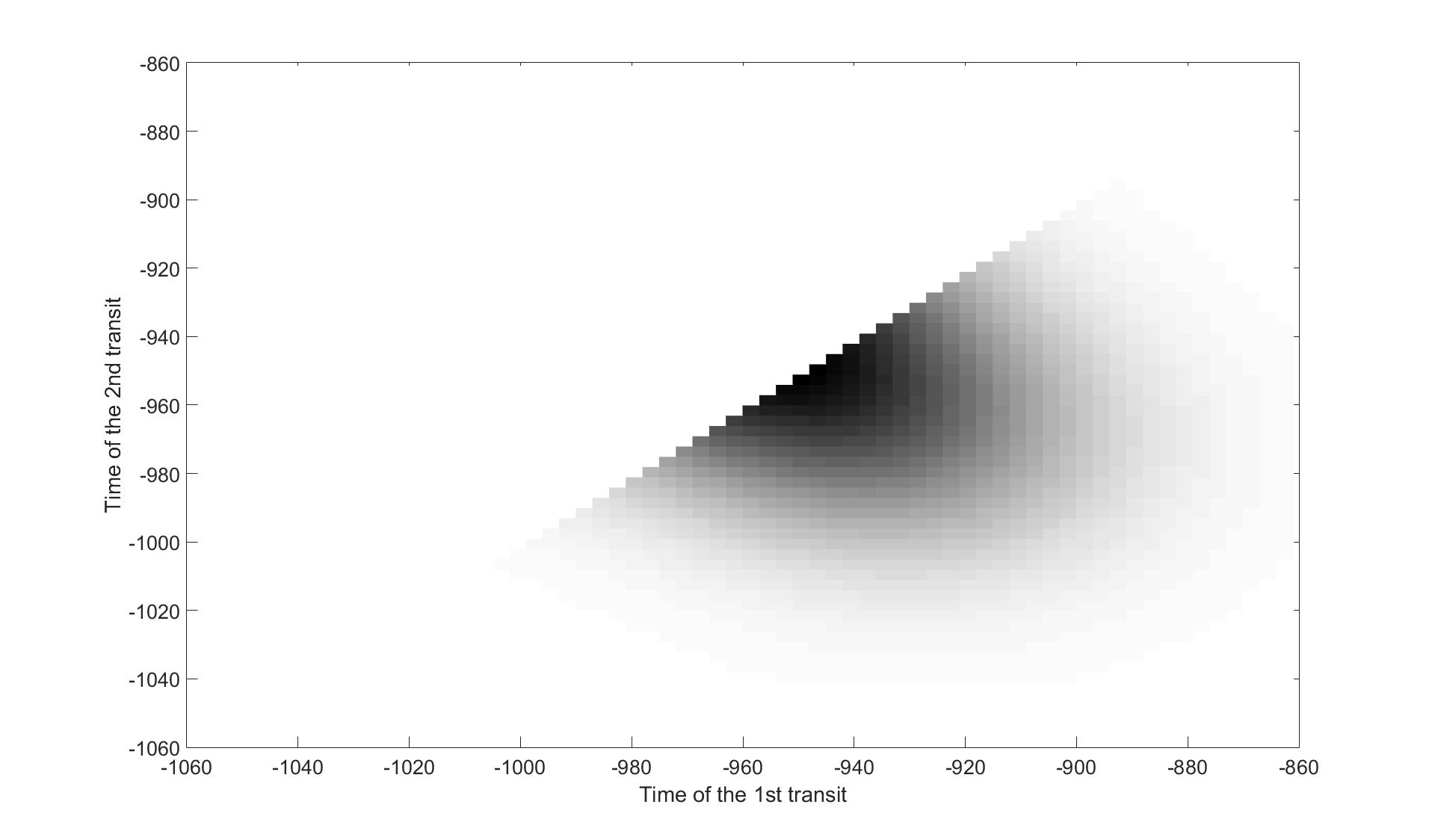}
\caption{Gray level map of the \emph{a posteriori} density of the two  transitions in a toy example.}
\label{fig:toyShortPeriod}
\end{center}
\end{figure}

The following simple toy example may explain this discrepancy. Suppose we had 4 observations belonging to 3
consecutive strata. The first strata has one observation at 980 BCE, the second has two observations equal to 980
and 920 BCE, and the third one observation at 920 BCE. The MLE is simple. The second strata is between 980 to 920
BCE. However, the Bayesian analysis is different. If we assume that the prior is uniform in the three strata,
flat for the transition times, and the all period is between 1250 to 650 BCE, then we obtain that the most likely
event is 0 length of the second strata. See Figure \ref{fig:toyShortPeriod}. More technically, the uniform prior
is more informative than it seems, as it puts strong preference for short periods mainly because of the
$\frac1{\tau_s-\tau_{s+1}}$ factors in \eqref{pss}. This can be corrected, for example by putting an informative prior on the
transition.  

Bootstraping of the Bayesian model is too computer intensive --- you need to calculate the \emph{a posteriori} at
any point for any bootstrap sample. Moreover, the bootstrap seems to be outside the basic philosophy of Bayesian analysis,
in particular, if the prior was honest, and in our case the statistical model is not regular to begin with. I avoided, therefore, doing the bootstrap.

The data I analyzed include a few sample points which seemingly should not influence the our understanding of
the transition times. Here I refer to samples R4, R5, R6, R12, R14, R15, R37, and R39. They are too early or too
late, to be informative on the start and end of their stratum respectively. Hence we may assume that taking them off, or adding
similar data points would not change the analysis. Similarly, the assumptions on the starting time of group I and
the end of group IV should not have an influence on the unrelated events (the end of group I and the beginning of
group IV)---they are too far apart.

\begin{figure}[t]
\begin{center}
\includegraphics[width=1\textwidth]{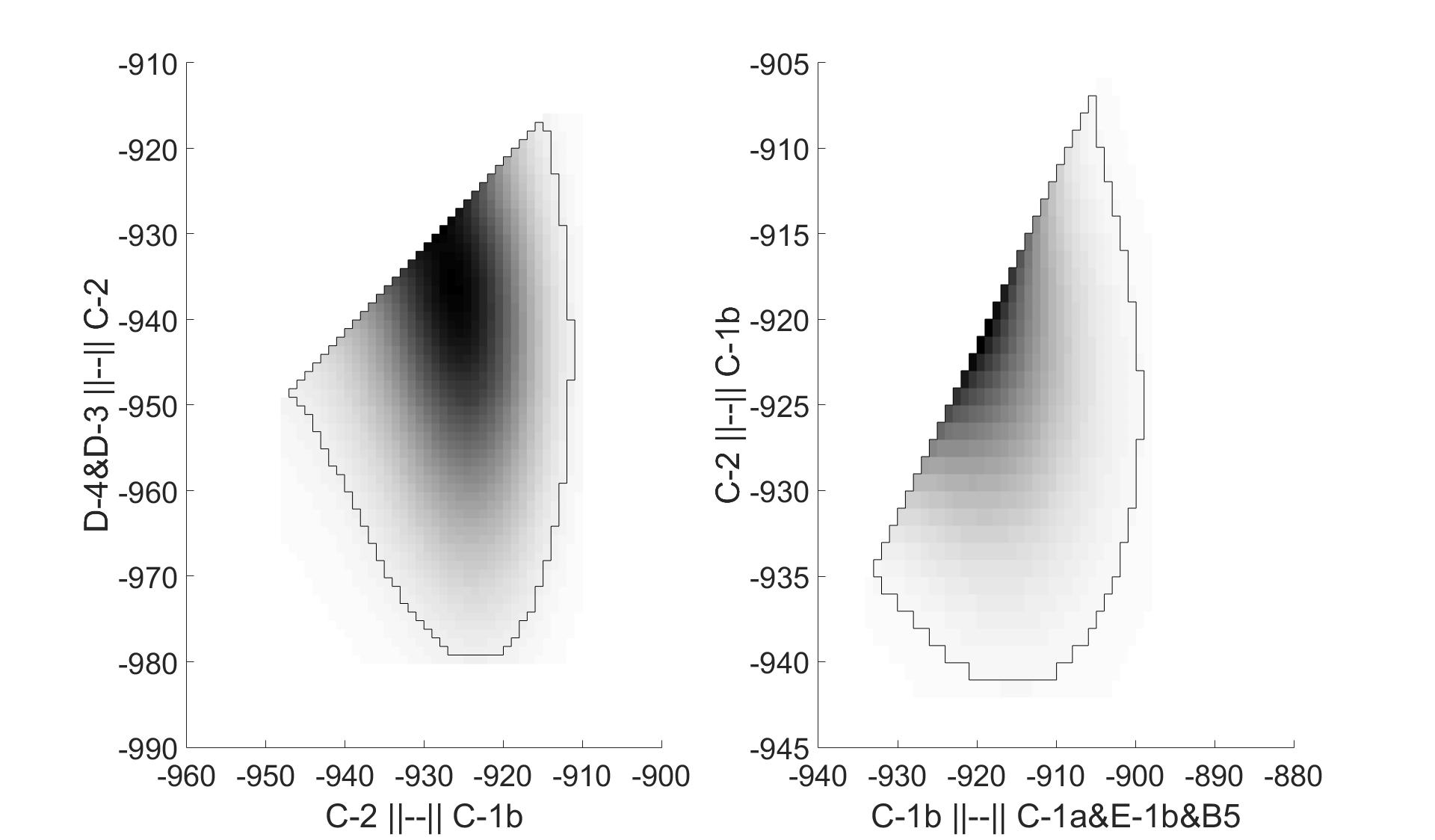}
\includegraphics[width=1\textwidth]{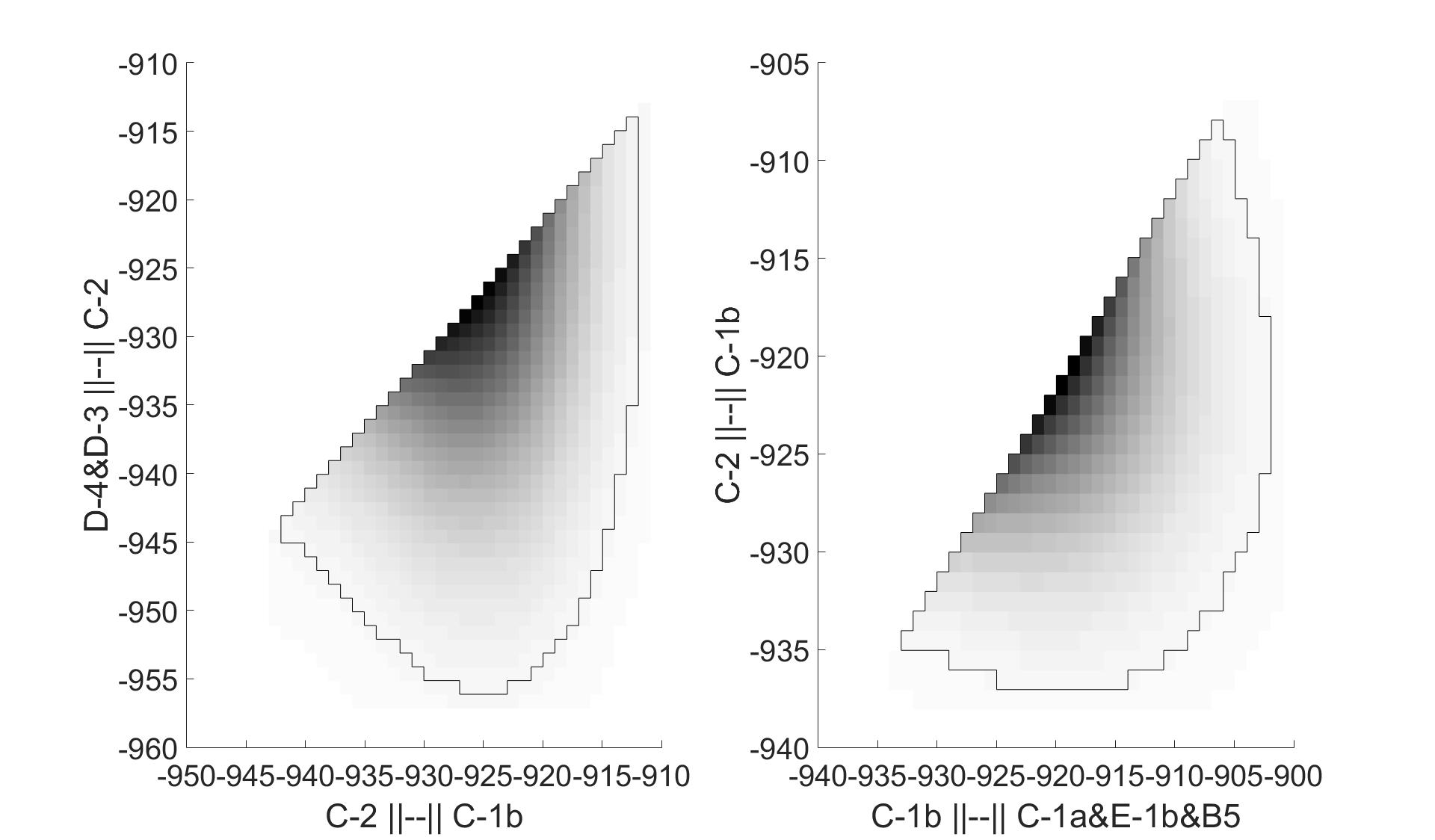}
\caption{Figure \ref{fig:bayesRehov}, but with multiple occurrences of remote samples (top), or without the
remote points (bottom).
The effect is not a result of the \emph{a priori} uniform distribution assumption. It would be the same whether
the density is triangular or even trapezoid. It will be whenever the density at the possible time of the remote
subsample is a function of the far away end point.}
\label{fig:bayesRehovMultiple}
\end{center}
\end{figure}

If we look on the expression of the \emph{a posteriori}, we obtain that this is not the case. The integrals over
these points are indeed almost independent of $\tau_1,\tau_2$, and $\tau_3$. However there is the factor
$\frac{1}{\tau_s-\tau_{s+1} }$ of \eqref{pss} discussed above. This factor is only due to the prior, and it favors short intervals.  To make the
situation extreme, I either multiplied each of these points  (i.e., to each of the above samples one similar
sample was added), or took it out altogether. Multiplying the observation favors short interval, and taking off the irrelevant observations makes long interval more likely. The results
are presented in Figure \ref{fig:bayesRehovMultiple}.

\section{Summary and conclusion}
\label{sec:summary}

The main purpose of this paper was a critical presentation of the Bayesian analysis of archeological data. We did it through a theoretical discussion of toy models, and a concrete analysis of specific data set --- the Iron I/IIA Age findings from Tel Rehov. Our conclusions from these two analyses is that the Bayesian approach is simple, intuitive, and natural. However, its conclusions depend on a
prior, that theoretically is supposed to be just that, what the archaeologist thought \emph{a priori} about the
values of the many different parameters of the problem, but \emph{de facto} it is what the programmer implemented
into the off the shelf program. A miscalculated prior can bias the analyze considerably.  I believe that robust priors --- priors that do not biased the analysis --- are conceptually inconsistent, and practically impossible for complex high dimensional and irregular models. The latter are the type of models that should be used for the analysis of Iron Age of the Levant.

The analysis was done on  restricted data, only those from Tel Rehov, and concrete conclusions for the scientific analysis of the period are beyond the scope of this paper. However, I believe that the range of the interpretation of the analyzed data was  presented in Figure \ref{fig:highLowWide}. Four consecutive strata were investigated. The first transit time seems to be any time in the third quarter of the tenth century BCE, while the third was either in the last quarter of the tenth century or the first of the ninth century BCE. The second stratum could have length from 0 to 20 years, the second and third strata could span together between twenty five to more than fifty years.

\section{Acknowledgment } I would like to thank Prof.\ Amihai Mazar who introduces me to this exciting field and
its challenges, and supply me with the Tel Rehov data. I would also want to acknowledge the contribution of an
enlightening conversation   with Ilan Sharon. This research was partially supported by ISF grant 1770/15.
\section{References}
\def\emph#1{#1}
\def\textbf#1{#1}

\begin{list}{}{\setlength{\itemindent}{-3ex}}
\item
Aczel AD 1995. Improved radiocarbon age estimating using the bootstrap.  \emph{Radiocarbon} \textbf{37}:845--849.
\item
Abramovich F,  Ritov Y  2013. Statistical Theory: A Concise Introduction. Boca Raton, London, New York: CRC Press,

\item Breiman L 1996. Bagging Predictors. \emph{Machine Learning}, 24:123--140.
\item
Bayliss A, Bronk Ramsey C 2004. Pragmatic Bayesians: a decade of integrating radiocarbon dates into
chronological models. In Buck CE and Millard AR editors. Constructing Chronologies: Crossing Disciplinary Boundaries, Lecture Notes in
Statistics,  177. London Springer-Verlag  p 25--41.
\item
Berger JO  1985. Statistical Decision Theory and Bayesian Analysis. New York Springer.
\item
Berger JO,  Berliner LM  1986. Robust Bayes and Empirical Bayes Analysis with \eps-Contaminated
Priors. Ann. Statist. 14:461--486.
\item
Bickel PJ, Doksum KA 2006. Mathematical Statistics, Basic Ideas and Selected Topics,
Vol. 1, (2nd Edition).  Upper Saddle River Prentice-Hall.
\item
Blackwell PG,  Buck CE 2008. Estimating radiocarbon calibration curves. Bayesian Analysis 3:225–48.
\item
Boaretto1 E, Timothy JAJ,  Gilboa A, Sharon I  2005.
Dating the Iron Age I/II transition in Israel: first inter-comparison. \emph{Radiocarbon},  \textbf{47}:39–-55.
\item
Bronk Ramsey C 2008. Radiocarbon Dating: Revolutions In Understanding. Archaeometry 50:249–275.
\item
Bronk Ramsey C 2009. Bayesian analysis of radiocarbon dates, Radiocarbon, 51:337--360.
\item
Bronk Ramsey C, Lee S 2013. Recent and planned developments of the program OxCal, Radiocarbon, 55:3--4.
\item
Bruins Hendrik J, van der Plicht J,  Mazar  A 2003. \c14  Dates from Tel Rehov: Iron-Age
Chronology, Pharaohs, and Hebrew Kings. Science 300:315--318.
\item Bruins  HJ,  van der Plicht J,  Mazar A,
Bronk Ramsey C, Manning SW  2005. The Groningen Radiocarbon
Series from Tel Rehov---
OxCal Bayesian computations for the Iron IB–
IIA Boundary and Iron IIA destruction events
In Levy TE  and  Highan T, editors. \emph{The Bible and Radiocarbon Dating}. London and Oakville Equinox.
\item
Buck CE, Christen JA,  James GN  1999. BCal: an on-line Bayesian radiocarbon calibration
tool, Internet Archaeology 7.
\item
Casella G  1985. An Introduction to Empirical Bayes Data Analysis. The American Statistician 39:83--87.
\item
 Finkelstein I 1995. The Date of the Settlement of the Philistines in Canaan. \emph{Tel Aviv} \textbf{22}:213--39.
Finkelstein I 1996. The Archaeology of the United Monarchy: An Alternative View. \emph{Levant} \textbf{28}:177--87.
\item
Gelman A, Carlin JB, Stern HS,   Rubin  D B 2014. Bayesian data analysis, volume 2. , New York Chapman and Hall.
\item
Geman S,  and Geman D  1984. Stochastic relaxation, Gibbs distributions, and the bayesian restoration
of images. IEEE Transactions on Pattern Analysis and Machine Intelligence 6:721--741.
\item
Godwin H  1962. Half-life of Radiocarbon. Nature 195:984.
\item
Hastings WK 1970. Monte carlo sampling methods using Markov chains and
their applications. Biometrika 57:97--109.
\item
Heaton TJ, Blackwell PG, Buck CE 2009. A Bayesian approach to the estimation of radiocarbon calibration
curves: the IntCal09 methodology. Radiocarbon 51:1151–64.
\item
Huber PJ  1964. Robust Estimation of a Location Parameter. The Annals of Mathematical Statistics 35:73--101.
\item
Huber PJ  2011. Robust statistics. New York Springer.
\item
Lee S , Bronk Ramsey C,  Mazar A  2013. Iron Age Chronology in Israel: Results from
Modeling with a Trapezoidal Bayesian Framework. Radiocarbon 55:731--740.
\item
Manning SW  2001. The absolute chronology of the Aegean early bronze age. Sheffield Sheffield Academic Press.
\item
Mazar A 2005. The debate over the chronology of the Iron Age in the southern Levant. In  Levy TE,   Highan T editors. \emph{The Bible and Radiocarbon Dating}.  London and Oakville  Equinox.
\item
Mazar, A 2013. Rehob. In Master D.M.,Nakhai B.  Alpert, Faust A, White L.M., and Zangeberg J.K.  editors. \emph{The Oxford Encyclopedia of Bible and Archaeology} \textbf{vol. 1}., Oxford University Press, New York.​
\item
Mazar A, Bruins HJ, Panitz-Cohen N, van der Plicht J. 2005. Ladder of time at Tel Rehov: stratigraphy,
archaeological context, pottery and radiocarbon dates. In Levy T, Higham T editors. The Bible and Radiocarbon
Dating: Archaeology, Text and Science. London Equinox p 195–255
\item
Mazar A  and Streit K  2016. Radiometric dates from Tel Reḥov (forthcoming).
\item
Niu M, Heaton TJ, Blackwell PG, Buck CE 2013. The Bayesian approach to radiocarbon calibration curve
estimation: the IntCal13, Marine13, and SHCal13 methodologies. Radiocarbon 55.
\item
Press SJ  2003. Subjective and Objective Bayesian Statistics 2nd edition. Hoboken Wiley.
\item
Reimer PJ, Bard E, Bayliss A, Beck JW, Blackwell PG,Bronk Ramsey C, Buck CE, Cheng H, Edwards RL, Friedrich M,
Grootes PM, Guilderson TP, Haflidason H, Hajdas I, Hatté C, Heaton TJ, Hoffman DL, Hogg AG, Hughen KA, Kaiser
KF, Kromer B, Manning SW, Niu M, Reimer RW, Richards DA, Scott EM, Southon JR, Staff RA, Turney CSM, van der
Plicht J (2013). IntCal13 and Marine13 radiocarbon age calibration curves 0–50,000 years cal BP. Radiocarbon
55.
\item
Ritov Y, Bickel PJ, Gamst  AC,  Kleijn BJK  2014. The Bayesian Analysis of Complex,
High-Dimensional Models: Can It Be CODA? \emph{Statist. Sci.}
29:619--639.
\item
Robert CP  2007. The Bayesian Choice From Decision-Theoretic Foundations to Computational
Implementation. New York Springer.
\item
Sharon I,  Gilboa A, Jull AJT,  Boaretto E 2007. Report on the first stage of the iron age dating project in Israel:
supporting a low chronology, \emph{Radiocarbon} \textbf{49}:1--46.
\item
Steier P  and Rom W  2000. The Use of Bayesian Statistics for \c14  Dates of Chronologically Ordered
Samples: A Critical Analysis. Radiocarbon,  42:183-–198.
\item
Stuiver M  and Polach HA 1977. Discussion: reporting of \c14  data. Radiocarbon, 19:355--363.
\item
 Tuo R,  Wu CFJ 2015. Efficient calibration for imperfect computer models. Annals of Statistics, to appear.
\item
van der Plicht J,  Bruins HJ 2005. Quality control of Groningen \c14 results from Tel Rehov. In Levy Thomas E.  and Highan Thomas editors. \emph{The Bible and Radiocarbon Dating}.  London and Oakville Equinox.
\end{list}

\end{document}